\documentclass[%
 reprint,
 amsmath,amssymb,
prb,
]{revtex4-1}

\usepackage{graphicx, dcolumn, bm, color}

\newcounter{RomanNumber}

\newcommand{\MC}[1]{\textcolor{black}{#1}}

\newcommand{\bfr}{\mathbf{r}}
\newcommand{\bfk}{\mathbf{k}}

\newcommand{\bfe}{\mathbf{e}}
\newcommand{\mcT}{\alpha}
\newcommand{\hpsi}{\widehat{\psi}}
\newcommand{\hphi}{\widehat{\phi}}
\newcommand{\tdta}{\mcT + \delta\mcT}
\newcommand{\tdtm}{\mcT - \delta\mcT}

\begin{document}

\title{Stability of three-dimensional icosahedral quasicrystals in multi-component systems}

\author{Kai Jiang*}
\affiliation{E-mail: kaijiang@xtu.edu.cn}

\author{Wei Si}

\affiliation{
 School of Mathematics and Computational Science,
 \\
Hunan Key Laboratory for Computation and Simulation in Science
and Engineering, Xiangtan University, P.R. China, 411105
}

\date{Jan 2019}

\begin{abstract}
	The relative stability of three-dimensional icosahedral
	quasicrystals in multi-component systems has been
	investigated based on a \MC{phenomenological} coupled-mode Swift-Hohenberg model with two-length-scales.  
	A recently developed projection method, which provides a unified numerical framework to study periodic crystals and quasicrystals, is used to compute free energy to high accuracy.
	Compared with traditional approaches, the advantage of the projection method has also been discussed \MC{in detail.}
	%#Compared with traditional approaches, the advantage of the projection method has also been discussed detailedly. 
    A rigorous and \MC{systematic} computation demonstrates that three-dimensional icosahedral quasicrystal \MC{and} two-dimensional decagonal quasicrystal are both stable phases in such a simple multi-component coupled-mode Swift-Hohenberg model. 
    %#A rigorous and systematical computation demonstrates that three-dimensional icosahedral quasicrystal, two-dimensional decagonal quasicrystal are both stable phases in such a simple multi-component coupled-mode Swift-Hohenberg model. 
	The result extends the two length-scales interaction mechanism of stabilizing quasicrystals from single-component to multi-component systems.
\begin{description}
\item[Keywords]
Icosahedral quasicrystals, Two-length-scales, Coupled-mode Swift-Hohenberg model,\\ Phase diagrams, Projection method
\item[PACS]61.44.Br, 61.44.Fw, 64.60.Bd, 02.70.Hm
\item[DOI]
\end{description}
\end{abstract}

\maketitle

\section{INTRODUCTION}
\label{sec.introduction}
Quasicrystals\,\cite{Levine1984} are a class of important ordered materials possessing quasiperiodic positional order and long-range orientational order between periodic crystals and amorphous materials.  
The mathematical description of quasicrystals can track back to the work done independently by Meyer\,\cite{Meyer1972} and Penrose\,\cite{penrose1974role} in early 1970. 
The first quasicrystal, actually a three dimensional (3D) icosahedral quasicrystal (IQC), was discovered by Shechtman in a rapidly-quenched Al-Mn alloy until 1982\,\cite{Shechtman1984}.
Since the first discovery of quasicrystals, the quasiperiodic long-range order has been found in a large number of metallic alloys\,\cite{Steurer2009, Steurer2004, Tsai2008}, and also in a host of soft-matter systems\,\cite{Zeng2004, Zhang2012}.  
Among these discoveries, icosahedral symmetric quasicrystals are the most frequently found, concretely in more than one hundred different metallic alloys\,\cite{Tsai2003, Tsai2008}.  
However, the thermodynamic stability of quasicrystals in multi-component systems, including IQCs, remains a challenge\,\cite{Henley2006, Lifshitz2007}, mainly due to the lack of appropriate theoretical models and high-precision numerical methods.

Theoretical approaches to investigating the stability of an ordered phase, including periodic and quasiperiodic one, often involve minimizing an appropriate free energy functional of the system and comparing the free energy of different candidate structures. 
A systematic examination of the relative stability of quasicrystals requires the availability of suitable free energy functionals and accurate methods to evaluate the free energy of phases with quasicrystalline orders. 
Nowadays, various coarse-grained free energy functionals have been proposed to investigate the phase behavior of physical systems, especially the thermodynamic stability of ordered phases including quasicrystals and periodic crystals\,\cite{Alexander1978, Bak1985, Jaric1985, Kalugin1985, Gronlund1988, Lifshitz1997, Subramanian2016, Jiang2017}.
Some of these works show the possibility of stable IQCs based on Landau-type free energy functionals with one order parameter\,\cite{Subramanian2016, Jiang2017}. 
Inspired by the \MC{Alexander-McTague} theory\,\cite{Alexander1978}, Mermin and Troian\,\cite{Mermin1985} introduced a second order parameter to obtain stable IQCs.  
%#Inspired by the Alexander-McTague's theory\,\cite{Alexander1978}, Mermin and Troian\,\cite{Mermin1985} introduced a second order parameter to obtain stable IQCs.  
On the other hand, Swift and Hohenberg\,\cite{Swift1977} explicitly added a positive-definite gradient term into the free energy functional to represent the effect of characteristic length-scales. 
This idea was rapidly extended to more than one characteristic length-scale and widely utilized to explore the quasicrystalline order\,\cite{Mermin1985, Mueller1994, Lifshitz1997, Dotera2007, Archer2013, Jiang2016}.
In particular, Dotera\,\cite{Dotera2007} extended \MC{the} Mermin-Troian model to study \MC{an} ABC star copolymer system with an incompressible condition.
%#In particular, Dotera\,\cite{Dotera2007} extended Mermin-Troian model to study ABC star copolymer system with an incompressible condition.
His work shows the emergence and stability of two-dimensional (2D) decagonal quasicrystals (DQCs). 
Recently, Jiang \textit{et al.}\,\cite{Jiang2016} have constructed a two-component \textit{coupled-mode Swift-Hohenberg} (CMSH) model with two length-scales to explore the thermodynamic stability of the periodic and quasiperiodic structures.
According to their work, the 2D decagonal and dodecagonal quasicrystals are both stable when the ratio of two-length-scales is appropriately chosen.
In the \MC{present} work, we will study the relative stability of 3D IQCs in multi-component systems based on the CMSH model.
%#In the recent work, we will study the relative stability of 3D IQCs in multi-component systems based on the CMSH model.

Besides a proper free energy functional to describe multi-component systems, examining the thermodynamic stability of quasicrystals requires accurate and efficient methods to evaluate the free energy of various ordered phases. 
Due to the \MC{lack of translational} symmetry, the computation of quasicrystals \MC{is harder to carry out} within a finite domain as done for periodic crystals. 
%#Due to the lack of the translational symmetry, the computation of quasicrystals is hardly carried out within a finite domain as done for periodic crystals. 
In the literature, utilizing a large periodic structure to approximate a quasicrystal is a commonly used method in the study of the quasicrystalline order\,\cite{Lifshitz1997, Subramanian2016, Quandt1999, Skibinsky1999, Dotera2006}. 
The method actually obtains a crystalline approximant, therefore, it is named the crystalline approximant method (CAM).
From the viewpoint of numerical computation, CAM is based on the approximation of irrational numbers by integers or rational numbers, corresponding to well-known \textit{Diophantine approximation} (DA) problem in the number theory\,\cite{Meyer1972, Jiang2014}. 
Because of the existence of DA, CAM has to be implemented in a very large computational region which means an unacceptable computational amount if small DA error is required.
Furthermore, it has been verified that the gap between the free energy of quasicrystals and their corresponding approximants cannot be vanished in any finite computational region\,\cite{Meyer1972}.  
In order to avoid the approximation error, an alternative approach is proposed to calculate quasicrystals based on the fact that quasiperiodic lattices can be generated by a cut-and-project method from higher-dimensional periodic lattices\,\cite{Meyer1972}.
This method provides a basic framework to study quasicrystals, originally proposed by Meyer in studying the relationship between harmonic analysis and algebraic numbers\,\cite{Meyer1972}.  
More recently, Jiang and Zhang proposed a projection method (PM) to obtain the density profile of quasicrystals and evaluate their energy density on high accuracy.
The PM shows that the Fourier spectrum of a quasiperiodic structure can be embedded into a higher-dimensional crystallographic point packing set of corresponding periodic structure.
Consequently, quasicrystals can be computed precisely in a higher-dimensional space and then be recovered by projecting the higher-dimensional reciprocal lattice vectors back to the original Fourier space through a projection matrix.
This method can avoid the DA error effectively.
As a particular case, it can be further used to investigate periodic crystals by setting the projection matrix as an identity matrix.  
From this perspective, the PM can calculate free energy of quasicrystals and periodic crystals with the same accuracy.
Therefore, the PM provides a unified computational framework to study the relative stability of quasicrystals and periodic crystals.

In the present work, we will mainly explore the formation and the thermodynamic stability of quasicrystals and periodic crystals in multi-component systems based on the CMSH model. 
A careful comparison of the free energy of abundant possible ordered phases, including 3D IQCs, 2D DQCs and periodic crystals, leads to confirm the relative stability and construct phase diagrams.
Theoretical results predict that IQCs can emerge and occupy a thermodynamic stable region in the phase diagram within the multi-component CMSH model.

\section{THEORETICAL FRAMEWORK}
\label{sec.theoretical}
Since the discovery of quasicrystals, a large number of theoretical studies have been carried out to investigate their symmetry, structure characterization physical properties and thermodynamic stability.
Current theoretical results demonstrate that the formation and stability of the quasicrystalline phases may be characterized by two or more length-scales interaction potential\,\cite{Mermin1985, Mueller1994, Lifshitz1997, Dotera2007, Archer2013, Jiang2016, Subramanian2016}.
Furthermore, since the formation of quasicrystals in multi-component systems could be hardly described by only one order parameter, more order parameters should be considered in a coarse-grained free energy functional.
Combining with the above factors, the CMSH model\,\cite{Jiang2016} was proposed to exploit the formation and the thermodynamic stability of quasicrystals.
In order to examine the relative stability of different ordered phases, the free energy of candidate structures, corresponding to local minima of the free energy functional, should be calculated precisely.
With an appropriate representation of quasicrystals, the PM provides a unified numerical framework to study periodic and quasiperiodic structures with the same precision\,\cite{Jiang2014}.
The rest of this section will give a brief introduction of the CMSH model and the PM for quasicrystals and periodic crystals.

\subsection{Coupled-mode Swift-Hohenberg model}
\label{subsec.model}
The original CMSH model only considers two-component systems with two-length-scales\,\cite{Jiang2016}. 
In this paper, we will give a general framework of multi-component systems with multi-length-scales. 
To be more concrete, the free energy functional of the CMSH model for a $m$-component system can be written as 
\begin{equation}
\begin{aligned}
    &F[\varphi_1, \varphi_2, \cdots, \varphi_m]
    \\
    =\;& \frac{1}{V}\Bigg\{ \frac{c}{2}\int 
    \sum_{j=1}^m [(\nabla^2 + q_j^2)\varphi_j(\mathbf{r})]^2
	+ \sum_{\mbox{\tiny$\begin{array}{c}i_j\in\{1,2,\cdots,m\} \\
		j=1,2,\cdots,n  \end{array}$ }}
    \\
    &\tau_{\sigma_{i_1}\sigma_{i_2}\cdots\sigma_{i_n} }
    [\sigma_{i_1}\varphi_{i_1}(\mathbf{r})]
	[\sigma_{i_2}\varphi_{i_2}(\mathbf{r})] \cdots
    [\sigma_{i_n}\varphi_{i_n}(\mathbf{r})]
	\,d\mathbf{r}
	\Bigg\} ,
\end{aligned}
	\label{eq:model}
\end{equation}
where $\varphi_j$ is the $j$-th order parameter.
$V$ represents the system volume.
$\sigma_{i_j}\in\{0,1\}$, $j=1,2,\cdots,n$. 
$q_j>0$ denotes the $j$-th characteristic length scale.
$c>0$ is an energy penalty factor to ensure $m$ characteristic length-scales.
The cross-terms in Eq.\eqref{eq:model} demonstrate the $n$-interactions among $m$ components. 
$ \tau_{\sigma_{i_1}\sigma_{i_2}\cdots\sigma_{i_n} }$ is interaction intensity related to physical conditions, such as temperature, pressure, and physical or chemical properties of materials.
Comparing with the common Landau theory of phase transition, the crucial feature of CMSH model is the occurrence of multi-length-scales $q_j$, $j=1,2,\cdots,m$. 
It should be noted that the number of characteristic multi-length-scales could be unequal to $m$.

In current work, we consider a two-component system with two-length-scales. 
At the same time, the highest-degree of interactions is fourth.
More specifically, the general CMSH model Eq.\eqref{eq:model} is simplified by the following free energy functional,
\begin{equation}
    \begin{aligned}
        F[\psi,\phi] =\; \frac{1}{V} \bigg\{ &\frac{c}{2} \int d\mathbf{r}
            [(\nabla^2+q_1^2)\psi]^2 + [(\nabla^2 + q_2^2)\phi]^2
            \\
            &+ \int d\mathbf{r} (\tau\psi^2 + g_0\psi^3 + \psi^4 + t\phi^2 + t_0\phi^3 
			\\
			&+ \phi^4 - g_1\psi^2\phi - g_2\psi\phi^2 + d_{0}\psi\phi
			\\
			&+ d_{1}\psi^{2}\phi^{2} + d_{2}\psi^{3}\phi + d_{3}\psi\phi^{3}) \bigg\}.
    \end{aligned}
    \label{eq.energy}
\end{equation}
Order parameters $\psi(\bfr)$ and $\phi(\bfr)$ correspond to the density profile of two-component systems such as metallic alloys or soft matters. 
$\tau$, $g_0$, $t$, $t_0$, $g_1$, $g_2$, $d_{0}$, $d_{1}$, $d_{2}$ and $d_{3}$ are all interaction parameters.
$q_j$, $j=1,2$ represents the characteristic length-scale.
According to the theories of Alexander and MacTague\,\cite{Alexander1978}, and Lifshitz and Petrich\,\cite{Lifshitz1997}, the cubic terms, associated with triangle interactions, in the above free energy functional play a crucial role in stabilizing periodic and quasiperiodic structures since they can reduce the value of free energy.
The ratio $q=q_2/q_1$ is related to the symmetry of ordered structures.
It should be pointed out that one of the length-scales can be always taken as a unit wavelength (as specified by $q_1=1$) and the second length-scale is specified by $q_2=q$.
The difference of the two-component CMSH model from a single-component incommensurate phase-field-crystal model, such as Lifshitz-Petrich model\,\cite{Lifshitz1997}, is the two-length-scales potential occurring in different order parameters.
Moreover, in consideration of an incompressibility condition\,\cite{Dotera2007, Jiang2016}, the two-component CMSH model can be utilized to describe three-component systems, such as Al-Cu-Fe alloys\,\cite{Tsai1987, Ishimasa1988}, Zn-Mg-Sc alloys\,\cite{Kaneko2001}, Al-Rh-Si alloys\,\cite{Koshikawa2003}, Au-Al-Yb alloys\,\cite{Ishimasa2011}, $etc$.
It should be remarked that the CMSH model is a \MC{phenomenological} model.
If the model parameters can match a given experimental system, then the CMSH model could be used to explain the corresponding physical system.
%#\MC{Because there are still too many variable model parameters in the model \eqref{eq.energyOriginal} to determine all possible cases, we only consider $ d_{0} = d_{1} = d_{2} = d_{3} = 0 $ to simplify the CMSH model in the following analysis and numerical computation.}
%#In the following analysis and numerical computation, we will set $ d_{0} = d_{1} = d_{2} = d_{3} = 0 $ to simplify the CMSH model.
%#The specific form is
%\begin{equation}
%    \begin{aligned}
%        F[\psi,\phi] =\; \frac{1}{V} \bigg\{ &\frac{c}{2} \int d\mathbf{r}
%            [(\nabla^2+q_1^2)\psi]^2 + [(\nabla^2 + q_2^2)\phi]^2
%            \\
%            &+ \int d\mathbf{r} (\tau\psi^2 + g_0\psi^3 + \psi^4 + t\phi^2 + t_0\phi^3 
%			\\
%			&+ \phi^4 - g_1\psi^2\phi - g_2\psi\phi^2 \bigg\}.
%    \end{aligned}
%    \label{eq.energy}
%\end{equation}

Theoretically, the ordered patterns including periodic and quasiperiodic structures correspond to local minima of the free energy functional with respect to order parameters $\psi$ and $\phi$ in the above energy functional.
Accordingly, the order parameters $\psi^{*}$ and $\phi^{*}$ located in the equilibrium state are the minima of the free energy density functional, which means
\begin{equation}
    \frac{\delta F}{\delta\psi(\mathbf{r})}\bigg|_{\psi^{*}} =\; 0,~~~~~~~~ 
    \frac{\delta F}{\delta\phi(\mathbf{r})}\bigg|_{\phi^{*}} =\; 0.
    \label{eq.condition}
\end{equation}
In order to find the equilibrium state, the Allen-Cahn dynamic equation is utilized to minimize the free energy functional and yields
\begin{equation}
    \left\{\;
    \begin{aligned}
        &\begin{aligned}
            \frac{\partial\psi}{\partial\mcT} =\; -\frac{\delta F}{\delta\psi}
            =\; &-c(\nabla^2+1)^2\psi-2\tau\psi-3g_0\psi^2
            \\
            &-4\psi^3+2g_1\psi\phi+g_2\phi^2 \MC{ - d_{0}\phi }
			\\
			& \MC{ - 2d_{1}\psi\phi^{2} - 3d_{2}\psi^{2}\phi - d_{3}\phi^{3} },
        \end{aligned}
        \\
        &\begin{aligned}
            \frac{\partial\phi}{\partial\mcT} =\; -\frac{\delta F}{\delta\phi} 
            =\; &-c(\nabla^2+q^2)^2\phi-2t\phi-3t_0\phi^2
            \\
            &-4\phi^3+2g_2\psi\phi+g_1\psi^2 \MC{ - d_{0}\psi }
			\\
			& \MC{ - 2d_{1}\psi^{2}\phi - d_{2}\psi^{3} - 3d_{3}\psi\phi^{2} }.
        \end{aligned}
    \end{aligned}
    \right.
    \label{eq.AC}
\end{equation}
The variable $ \mcT $ does not represent time but a parameter controlling the iteration steps.

\subsection{Projection Method}
\label{subsec.projection}
Due to the spatial periodicity, the computation of periodic crystals can be carried out within a unit cell with periodic boundary conditions.
On the contrary, this method is not suitable for quasicrystals because they are quasiperiodic in at least one direction.
For the quasicrystalline order, an efficient method is \textit{the projection method }(PM) based on the fact that a $d$-dimensional quasicrystal is a combination of a class of exponentials $\big\{e^{i\mathbf{k}^{T}\mathbf{r}}\big\}$, $\mathbf{r}\in\mathbb{R}^d$, $\mathbf{k}$ belongs to a countable set\,\cite{Jiang2014}.
From the higher-dimensional description\,\cite{Steurer2009}, any $d$-dimensional quasicrystal can be embedded into an $n$-dimensional periodic structure ($n \geqslant d$).
Specifically, the $n$-dimensional reciprocal vectors can be spanned by a set of primitive vectors $\mathbf{b}_i$, which are the primitive reciprocal vectors in the $n$-dimensional reciprocal space, with integer coefficients. 
It means any reciprocal vector of an $n$-dimensional periodic structure can be expressed as $\mathbf{H} = \mathbf{Bh}$, where $\mathbf{B} = (\mathbf{b}_1,\dots,\mathbf{b}_n)\in\mathbb{R}^{n \times n}$ is the primitive reciprocal lattice and $\mathbf{h}\in\mathbb{Z}^n$. 
Then the wave vector $\mathbf{k}$ in $d$-dimensional physical space is obtained by the projection, $\mathbf{k} = \mathcal{P}\cdot\mathbf{H}$, where $\mathcal{P}$ is a projection matrix of $d \times n$-order. 
The dimensionality $n$ is determined by the rotational symmetry of the quasicrystal. 
In particular, it can be obtained by an additive Euler function\,\cite{hiller1985crystallographic}.
In the view of numerical computation, the projection matrix can be directly confirmed through the representation of the primitive basis vectors\,\cite{Jiang2014}.

In detail, using the $n$-dimensional periodic structure and the projection matrix, any $d$-dimensional quasiperiodic function $\psi(\mathbf{r})$ can be expanded as
\begin{equation}
    \psi(\mathbf{r}) = \sum_{\mathbf{h}\in\mathbb{Z}^n} \widehat{\psi}(\mathbf{h})
    e^{i[(\mathcal{P}\cdot\mathbf{Bh})^{T}\cdot\mathbf{r}]},
    ~\mathbf{r}\in\mathbb{R}^d,
    \label{eq.expansion}
\end{equation}
where the Fourier coefficient $\widehat{\psi}(\mathbf{h})$ can be easily obtained by using the $n$-dimensional $L^2$-inner product, $\widehat{\psi}(\mathbf{h}) = \left<\tilde{\psi}(\tilde{\mathbf{r}}), e^{i[(\mathcal{P}\cdot\mathbf{Bh})^{T}\cdot\tilde{\mathbf{r}}]}\right>$, with $\{\widehat{\psi}(\mathbf{h})\}_{\mathbf{h}\in\mathbb{Z}^n} \in \ell^2(\mathbb{Z}^n)$ and $\tilde{\mathbf{r}} = \sum_{i=1}^n c_i \mathbf{a}_i, c\in[0,1)$.
$\mathbf{a}_i\in\mathbb{R}^n$, $i = 1,2,\dots,n$, is the reciprocal primitive vector which satisfies the dual relationship, $\mathbf{a}_i\cdot\mathbf{b}_j = 2\pi\delta_{ij}$. 
Furthermore, the function $\tilde{\psi}(\tilde{\mathbf{r}})$ is the inverse Fourier transform of the Fourier coefficient $\widehat{\psi}(\mathbf{h})$. 
From the expansion Eq.\eqref{eq.expansion}, the $d$-dimensional quasiperiodic structure can be also treated as a hyperplane of an $n$-dimensional periodic structure whose orientation is determined by the projection matrix $\mathcal{P}$. 
In order to describe the position of the quasilattice in $d$-dimensional Fourier space, the notation $\mathbf{k}$ is utilized to replace $\mathcal{P}\cdot\mathbf{Bh}$ in Eq.\eqref{eq.expansion}.
With this notation, the projection method has the following form,
\begin{equation}
    \psi(\mathbf{r}) = \sum_{\mathbf{k}}
	\widehat{\psi}(\mathbf{k})
    e^{i\mathbf{k}^{T}\cdot\mathbf{r}},
    ~\mathbf{r}\in\mathbb{R}^d,
    \label{eq.expan.psi}
\end{equation}
where $\mathbf{k} = \sum_{i=1}^n h_i(\mathcal{P}\mathbf{b}_i) \in \mathbb{R}^d$.
Despite being similar to the common Fourier series, it should be emphasized that the distribution of $\mathbf{k}$ is not a periodic lattice. 
Similarly, the order parameter $\phi(\mathbf{r})$ in multi-component systems can be expanded as follows,
\begin{equation}
    \phi(\mathbf{r}) = \sum_{\mathbf{k}}
	\widehat{\phi}(\mathbf{k})
    e^{i\mathbf{k}^{T}\cdot\mathbf{r}},
    ~\mathbf{r}\in\mathbb{R}^d,
    \label{eq.expan.phi}
\end{equation}
where $\mathbf{k}$ has the same definition as Eq.\eqref{eq.expan.psi}.

For a given structure of interest, the reciprocal lattice vectors are determined by its symmetry, and the optimal coefficients are obtained by minimizing the free energy functional.
Based on these expansions, we can compute quasicrystals in the high-dimensional space and then project it onto the lower-dimensional physical space. 
Since the reciprocal lattice in the $n$-dimensional space is periodic, the computation can be performed on a uniform mesh grid. 
As a particular case, a $d$-dimensional periodic structure can be described by the PM when the projection matrix is set as a $d$-order identity matrix.
In this sense, the PM becomes the common Fourier-spectral method.
Therefore, this perspective provides a unified computational framework of periodic and quasiperiodic structures with the same accuracy.
More significantly, the PM can evaluate the free energy of the corresponding ordered structure to high accuracy.

In practice, inserting the generalized Fourier expansions Eqs.\eqref{eq.expan.psi} and \eqref{eq.expan.phi} into the dynamic equations \eqref{eq.AC}, we can obtain the following iterative equations expressed by Fourier coefficients,
\begin{equation}
    \left\{\;
    \begin{aligned}
        &\begin{aligned}
            \frac{\partial\hpsi(\bfk)}{\partial\mcT} = 
            &-c(-|\bfk|^2+1)^2\hpsi(\bfk) - 2\tau\hpsi(\bfk) - 3g_0\widehat{\psi^2}(\bfk)
            \\
            &- 4\widehat{\psi^3}(\bfk) + 2g_1\widehat{(\psi\phi)}(\bfk) + g_2\widehat{\phi^2}(\bfk) \MC{ - d_{0}\widehat{\phi}(\bfk) }
			\\
			& \MC{ - 2d_{1}\widehat{(\psi\phi^{2})}(\bfk) - 3d_{2}\widehat{(\psi^{2}\phi)}(\bfk) - d_{3}\widehat{\phi^{3}}(\bfk) },
        \end{aligned}
        \\
        &\begin{aligned}
            \frac{\partial\hphi(\bfk)}{\partial\mcT} = 
            &-c(-|\bfk|^2+q^2)^2\hphi(\bfk) - 2t\hphi(\bfk) - 3t_0\widehat{\phi^2}(\bfk) 
            \\
            &- 4\widehat{\phi^3}(\bfk) + 2g_2\widehat{(\psi\phi)}(\bfk) + g_1\widehat{\psi^2}(\bfk) \MC{ - d_{0}\widehat{\psi}(\bfk) }
			\\
			& \MC{ - 2d_{1}\widehat{(\psi^{2}\phi)}(\bfk) - d_{2}\widehat{\psi^{3}}(\bfk) - 3d_{3}\widehat{(\psi\phi^{2})}(\bfk) }.
        \end{aligned}
    \end{aligned}
    \right.
    \label{eq.AC.fft}
\end{equation}
In this expression, the quadratic, cubic and cross terms are given by,
\begin{equation}
    \left\{\;
    \begin{aligned}
        &\widehat{\psi^2}(\bfk) = \sum_{\mbox{\tiny$\begin{array}{c}
                    |\bfk_1|=|\bfk_2|=1\\
                    \bfk_1+\bfk_2=\bfk\end{array}$}}
            \hpsi(\bfk_1)\hpsi(\bfk_2),
        \\
        &\widehat{\phi^2}(\bfk) = \sum_{\mbox{\tiny$\begin{array}{c}
                    |\bfk_1|=|\bfk_2|=q\\
                    \bfk_1+\bfk_2=\bfk\end{array}$}}
            \hphi(\bfk_1)\hphi(\bfk_2),
        \\
        &\widehat{\psi^3}(\bfk) = \sum_{\mbox{\tiny$\begin{array}{c}
                    |\bfk_1|=|\bfk_2|=|\bfk_3|=1\\
                    \bfk_1+\bfk_2+\bfk_3=\bfk\end{array}$}}
            \hpsi(\bfk_1)\hpsi(\bfk_2)\hpsi(\bfk_3),
        \\
        &\widehat{\phi^3}(\bfk) = \sum_{\mbox{\tiny$\begin{array}{c}
                    |\bfk_1|=|\bfk_2|=|\bfk_3|=q\\
                    \bfk_1+\bfk_2+\bfk_3=\bfk\end{array}$}}
            \hphi(\bfk_1)\hphi(\bfk_2)\hphi(\bfk_3),
        \\
        &\widehat{(\psi\phi)}(\bfk) = \sum_{\mbox{\tiny$\begin{array}{c}
                    |\bfk_1|=1,|\bfk_2|=q\\
                    \bfk_1+\bfk_2=\bfk\end{array}$}}
            \hpsi(\bfk_1)\hphi(\bfk_2)
        \\
        & \MC{ \widehat{(\psi\phi^{2})}(\bfk) = \sum_{\mbox{\tiny$\begin{array}{c}
                    |\bfk_1|=1,|\bfk_2|=|\bfk_3|=q\\
                    \bfk_1+\bfk_2+\bfk_3=\bfk\end{array}$}}
            \hpsi(\bfk_1)\hphi(\bfk_2)\hphi(\bfk_3) }
        \\
        & \MC{ \widehat{(\psi^{2}\phi)}(\bfk) = \sum_{\mbox{\tiny$\begin{array}{c}
                    |\bfk_1|=|\bfk_2|=1,|\bfk_3|=q\\
                    \bfk_1+\bfk_2+\bfk_3=\bfk\end{array}$}}
            \hpsi(\bfk_1)\hpsi(\bfk_2)\hphi(\bfk_3) }.
    \end{aligned}
    \right.
    \label{eq.expan.psi.phi}
\end{equation}
From these expressions, it is clear that the nonlinear (quadratic, cubic and cross) terms in Eq.\eqref{eq.AC.fft} are $n$-dimensional convolutions in the reciprocal space. 
A direct evaluation of these nonlinear terms is extremely expensive. 
Instead, these terms are simple multiplication in the $n$-dimensional real space.
The pseudospectral method takes advantage of this observation by evaluating the gradient terms in the Fourier space and the nonlinear terms in the real space by performing the efficient Fast Fourier Transformation (FFT) algorithm.  
Thus it provides an efficient technique to find the solutions of dynamic equations.

In order to solve the time-dependent equations \eqref{eq.AC.fft} numerically, it is necessary to apply a time discretization scheme.
In this work, we propose a second-order precision method by combining the second-order Adam-Bashforth with Lagrange extrapolation approach (BDF2-LE) to \MC{discretize} the dynamic equations.
%#In this work, we propose a second-order precision method by combining the second-order Adam-Bashforth with Lagrange extrapolation approach (BDF2-LE) to discrete the dynamic equations.
In particular, the scheme can be written as
\begin{equation}
    \left\{\;
    \begin{aligned}
        &\begin{aligned}
            (\hpsi)_{\text{BDF2}} =\;
            &-c(-|\bfk|^2+1)^2\hpsi_{\tdta} - 2\tau\hpsi_{\tdta}
			\\
			&- 3g_0\widehat{\bar{\psi}^2}_{\tdta} - 4\widehat{\bar{\psi}^3}_{\tdta} + 2g_1\widehat{(\bar{\psi}\bar{\phi})}_{\tdta}
			\\
			& + g_2\widehat{\bar{\phi}^2}_{\tdta} \MC{ - d_{0}\widehat{\bar{\phi}}_{\tdta} - 2d_{1}\widehat{(\bar{\psi}\bar{\phi}^{2})}_{\tdta} }
			\\
			& \MC{ - 3d_{2}\widehat{(\bar{\psi}^{2}\bar{\phi})}_{\tdta} - d_{3}\widehat{\bar{\phi}^{3}}_{\tdta} }, 
        \end{aligned}
        \\
        &\begin{aligned}
            (\hphi)_{\text{BDF2}} =\;
            &-c(-|\bfk|^2+q^2)^2\hphi_{\tdta} - 2t\hphi_{\tdta}
			\\
			&- 3t_0\widehat{\bar{\phi}^2}_{\tdta} - 4\widehat{\bar{\phi}^3}_{\tdta} + 2g_2 \widehat{(\bar{\psi}\bar{\phi})}_{\tdta}
			\\
			& + g_1\widehat{\bar{\psi}^2}_{\tdta} \MC{ - d_{0}\widehat{\bar{\psi}}_{\tdta} - 2d_{1}\widehat{(\bar{\psi}^{2}\bar{\phi})}_{\tdta} }
			\\
			& \MC{ - d_{2}\widehat{\bar{\psi}^{3}}_{\tdta} - 3d_{3}\widehat{(\bar{\psi}\bar{\phi}^{2})}_{\tdta} }, 
        \end{aligned}
		\\
		& (\bar{\psi}\bar{\phi})_{\tdta} = \bar{\psi}_{\tdta} \bar{\phi}_{\tdta},
		\\
		& (\bar{\psi}^{2}\bar{\phi})_{\tdta} = \left(\bar{\psi}_{\tdta}\right)^{2} \bar{\phi}_{\tdta},
		\\
		& (\bar{\psi}\bar{\phi}^{2})_{\tdta} = \bar{\psi}_{\tdta} \left(\bar{\phi}_{\tdta}\right)^{2},
    \end{aligned}
    \right.
    \label{eq.BDF2}
\end{equation}
where $ (\cdot)_{\text{BDF2}} = \dfrac{3(\cdot)_{\tdta} - 4(\cdot)_{\mcT}  + (\cdot)_{\tdtm}}{2\delta\mcT} $, and $ \bar{(\cdot)}_{\tdta} $ can be calculated by the Lagrange extrapolation approach $ \bar{(\cdot)}_{\tdta} = 2(\cdot)_{\mcT} - (\cdot)_{\tdtm} $.

\subsection{Two Modes Approximation Method}
\label{subsec.TMA}

From the observation of the CMSH model \eqref{eq.energy}, it is clear that the penalty factor $c$ affects the emergence of nonzero Fourier vectors. 
In order to systematically investigate the ideal and actual phase behavior, we consider the hard constraint ($c\to+\infty$) and the soft constraint (finite $c$).  
Under the hard constraint, $\hpsi$ (resp.~$\hphi$) should be strictly restricted on the circle with radius $1$ (resp.~$q$), otherwise, the value of the free energy functional \eqref{eq.energy} will be infinite.
Therefore, it is only required to analyze the entropy part.
This is the key idea of the two modes approximation method (TMAM).
In particular, \MC{under} the hard constraint, the expansion terms of Eqs.\eqref{eq.expan.psi} and \eqref{eq.expan.phi} are finite for any given symmetric structure.
%#In particular, Under the hard constraint, the expansion terms of Eqs.\eqref{eq.expan.psi} and \eqref{eq.expan.phi} are finite for any given symmetric structure.
Inserting Eqs.\eqref{eq.expan.psi} and \eqref{eq.expan.phi} into the CMSH model \eqref{eq.energy}, we obtain the following expression with respect to Fourier coefficients $\hpsi(\bfk)$ and $\hphi(\bfk)$,
\begin{equation}
	\begin{aligned}
		F&[\hpsi, \hphi] =\;
		\tau\sum_{\mbox{\tiny$\begin{array}{c}
			|\bfk_1|=|\bfk_2|=1\\
			\bfk_{1}+\bfk_{2}=\mathbf{0}\end{array}$}}
		\hpsi(\bfk_{1})\hpsi(\bfk_{2})
		\\
		&+ g_0\sum_{\mbox{\tiny$\begin{array}{c}
			|\bfk_1|=|\bfk_2|=|\bfk_3|=1\\
			\bfk_{1}+\bfk_{2}+\bfk_{3}=\mathbf{0}\end{array}$}}
		\hpsi(\bfk_1)\hpsi(\bfk_2)\hpsi(\bfk_3)
		\\
		&+ \sum_{\mbox{\tiny$\begin{array}{c}
			|\bfk_1|=|\bfk_2|=|\bfk_3|=|\bfk_4|=1\\
			\bfk_{1}+\bfk_{2}+\bfk_{3}+\bfk_{4}=\mathbf{0}\end{array}$}}
		\hpsi(\bfk_1)\hpsi(\bfk_2)\hpsi(\bfk_3)\hpsi(\bfk_4)
		\\
		&+ t\sum_{\mbox{\tiny$\begin{array}{c}
			|\bfk_1|=|\bfk_2|=q\\
			\bfk_{1}+\bfk_{2}=\mathbf{0}\end{array}$}}
		\hphi(\bfk_1)\hphi(\bfk_2)
		\\
		&+ t_0\sum_{\mbox{\tiny$\begin{array}{c}
			|\bfk_1|=|\bfk_2|=|\bfk_3|=q\\
			\bfk_{1}+\bfk_{2}+\bfk_{3}=\mathbf{0}\end{array}$}}
		\hphi(\bfk_1)\hphi(\bfk_2)\hphi(\bfk_3)
		\\
		&+ \sum_{\mbox{\tiny$\begin{array}{c}
			|\bfk_1|=|\bfk_2|=|\bfk_3|=|\bfk_4|=q\\
			\bfk_{1}+\bfk_{2}+\bfk_{3}+\bfk_{4}=\mathbf{0}\end{array}$}}
		\hphi(\bfk_1)\hphi(\bfk_2)\hphi(\bfk_3)\hphi(\bfk_4)
		\\
		&- g_1\sum_{\mbox{\tiny$\begin{array}{c}
			|\bfk_1|=|\bfk_2|=1,|\bfk_{3}|=q\\
			\bfk_{1}+\bfk_{2}+\bfk_{3}=\mathbf{0}\end{array}$}}
		\hpsi(\bfk_1)\hpsi(\bfk_2)\hphi(\bfk_3)
		\\
		&- g_2\sum_{\mbox{\tiny$\begin{array}{c}
			|\bfk_1|=1, |\bfk_2|=|\bfk_{3}|=q\\
			\bfk_{1}+\bfk_{2}+\bfk_{3}=\mathbf{0}\end{array}$}}
		\hpsi(\bfk_1)\hphi(\bfk_2)\hphi(\bfk_3)
		\\
		& \MC{ + d_{0} \sum_{\mbox{\tiny$\begin{array}{c}
			|\bfk_1|=1, |\bfk_2|=q\\
			\bfk_{1}+\bfk_{2}=\mathbf{0}\end{array}$}}
		\hpsi(\bfk_1)\hphi(\bfk_2) }
		\\
		& \MC{ + d_{1} \sum_{\mbox{\tiny$\begin{array}{c}
			|\bfk_1|=|\bfk_{2}|=1, |\bfk_{3}|=|\bfk_{4}|=q\\
			\bfk_{1}+\bfk_{2}+\bfk_{3}+\bfk_{4}=\mathbf{0}\end{array}$}}
		\hpsi(\bfk_1)\hpsi(\bfk_{2})\hphi(\bfk_{3})\hphi(\bfk_{4}) }
		\\
		& \MC{ + d_{2} \sum_{\mbox{\tiny$\begin{array}{c}
			|\bfk_1|=|\bfk_{2}|=|\bfk_{3}|=1, |\bfk_{4}|=q\\
			\bfk_{1}+\bfk_{2}+\bfk_{3}+\bfk_{4}=\mathbf{0}\end{array}$}}
		\hpsi(\bfk_1)\hpsi(\bfk_{2})\hpsi(\bfk_{3})\hphi(\bfk_{4}) }
		\\
		& \MC{ + d_{3} \sum_{\mbox{\tiny$\begin{array}{c}
			|\bfk_1|=1, |\bfk_{2}|=|\bfk_{3}|=|\bfk_{4}|=q\\
			\bfk_{1}+\bfk_{2}+\bfk_{3}+\bfk_{4}=\mathbf{0}\end{array}$}}
		\hpsi(\bfk_1)\hphi(\bfk_{2})\hphi(\bfk_{3})\hphi(\bfk_{4}) }.
	\end{aligned}
	\label{eq.energy.k}
\end{equation}
It should be noted that the summations are all finite in the above expression.
Through the TMAM, the free energy functional is turned into a function \MC{with a finite number of variables}.
%#Through the TMAM, the free energy functional is turned into a function with finite variables.
The energy expressions of all involved ordered structures can be found in Sec.\,\ref{subsec.results}.

\section{RESULTS AND DISCUSSION}
\label{sec.results}
In this section, all experiments were performed on a desktop computer with a 3.20 GHz CPU (i5-6500, 4 processors).
All codes were written \MC{in MATLAB without parallelization}.
%#All codes were written by MATLAB language without parallel.
%#\MC{Because there are too many variable model parameters in the model \eqref{eq.energy} to determine all possible cases, we only consider $ d_{0} = d_{1} = d_{2} = d_{3} = 0 $ as an example to investigate the stability of 3D IQCs in the following analysis and numerical computation.}

\subsection{The advantages of PM}
\label{subsec.CAM.PM}
High accurate numerical methods are crucial to theoretically investigating the relative stability of ordered structures.
The CAM is a widely used approach to examine quasiperiodic structures, however, it inevitably accompanies the DA error, $E_{DA}$\,\cite{Meyer1972, Jiang2014}, which dominates the computational precision.
In order to avoid $E_{DA}$, the PM is a reliable approach\,\cite{ Jiang2014}.
In this subsection, we will demonstrate that the PM has advantages over the CAM.
The main idea of CAM is using a large periodic structure to approximate a quasicrystal. 
For a desired computational precision, CAM always requires a large computational region to reduce $E_{DA}$.
More details about CAM can be found in Appendix \ref{app.introd.CAM}.
Recently, based on the observation that a quasiperiodic phase can be embedded in a higher-dimensional periodic structure, an efficient method, \textit{i.e.}, PM, has been formulated in the Fourier space. 
Comparing with CAM, PM not only requires less computational cost, but also computes real quasicrystalline structures, as well as their free energy to high accuracy.

In the following, we will take the 2D DQC as an example to demonstrate the correctness and efficiency of PM in detail, the morphology of 2D DQC can be found in FIG.\,\ref{fig.can.real} (a).  
The 2D DQC can be embedded into a $4$-dimensional periodic structure.
Therefore the calculation of PM is implemented in 4D space while that of CAM is carried out in 2D space.
In order to compare the computational cost of both approaches, we set $N$ discrete points in the length $2\pi$ of each direction.
In practice, the computational region of PM is $[0,2\pi)^4$ in 4D space, but that of CAM should be $[0, L\cdot 2\pi)^2$, where the value of $L\in\mathbb{Z}$ is dependent on the desired precision of $E_{DA}$.
Therefore the number of discrete points of PM and CAM is $ N^4 $ and $ (LN)^2 $, respectively.
As FIG.\,\ref{fig.DA} (in Appendix \ref{app.introd.CAM}) shows, $E_{DA}$ does not \MC{decay monotonically} as $L$ increases. 
%#As FIG.\,\ref{fig.DA} (in Appendix \ref{app.introd.CAM}) shows, $E_{DA}$ does not monotonously decay as $L$ increases. 
The minimal integer $L$ of desired $E_{DA}$ for 2D DQC is listed in TAB.\,\ref{tab.DA}. 
\begin{table}[htbp]
	\centering
	\caption{\label{tab.DA}
	The minimal integer $L$ for desired $E_{DA}$.
	The size of computational domain is $[0,L\cdot 2\pi)^2$ in CAM.
	}
	\begin{tabular}{|c|c|c|c|c|c|}
		\hline
		$E_{DA}$ & 0.166879 & 0.091809 & 0.067405 & 0.055535 & 0.037427\\
		\hline
		$L$ & 126 & 204 & 1288 & 2084 & 3372\\
		\hline
	\end{tabular}
\end{table}
It should be pointed out that the PM computes real DQCs, while the CAM merely calculates corresponding crystalline approximants.
\begin{table}[htbp]
	\centering
	\caption{\label{tab.CPU}
		A comparison of the convergent free energy calculated by PM and CAM with different numbers of discrete points $N$.
		The model parameters set as $c=80,\tau=0,t=0,t_0=-0.3,g_0=-0.3,g_1=2.2, g_2=2.2, d_{0} = d_{1} = d_{2} = d_{3} = 0 $.
		}
	\begin{tabular}{|c|c|c|}
		\hline
		$N$ & PM & CAM($L=126$)\\
		\hline
		10 & -4.296833632029e-02 & -4.287269226697e-02\\
		\hline
		12 & -4.296833659056e-02 & -4.287269226627e-02\\
		\hline
		14 & -4.296833659212e-02 & -4.287269226626e-02\\
		\hline
		16 & -4.296833659224e-02 & -4.287269226626e-02\\
		\hline
		18 & -4.296833659225e-02 & -4.287269226626e-02\\
		\hline
		20 & -4.296833659225e-02 & -4.287269226626e-02\\
		\hline
		22 & -4.296833659225e-02 & -4.287269226626e-02\\
		\hline
	\end{tabular}
\end{table}
TAB.\,\ref{tab.CPU} gives the free energy of DQCs calculated by PM and by CAM ($L=126$, $E_{DA}=0.166879$) \MC{with different discretizations} when $ c=80 $, $ \tau=0 $, $ t=0 $, $ t_0=-0.3 $, $ g_0=-0.3 $, $ g_1=2.2 $, \MC{$ g_2=2.2 $ and $ d_{0} = d_{1} = d_{2} = d_{3} = 0 $}.
%#TAB.\,\ref{tab.CPU} gives the free energy of DQCs calculated by PM and by CAM ($L=126$, $E_{DA}=0.166879$) with different discretized points when $ c=80 $, $ \tau=0 $, $ t=0 $, $ t_0=-0.3 $, $ g_0=-0.3 $, $ g_1=2.2 $ and $ g_2=2.2 $.
Applying CAM to compute the crystalline approximants of DQCs, the minimum computational region is $ [0, 126\times 2\pi)^2 $.
It \MC{was} found that two approaches obtain different convergent energy values as $ N $ increases.
%#It could be found that two approaches obtain different convergent energy values as $ N $ increases.
While $ N \geq 18 $, the free energy computed by both methods has $13$ significant digits.

\begin{figure}[htbp]
	\centering
	\includegraphics[scale=0.15]{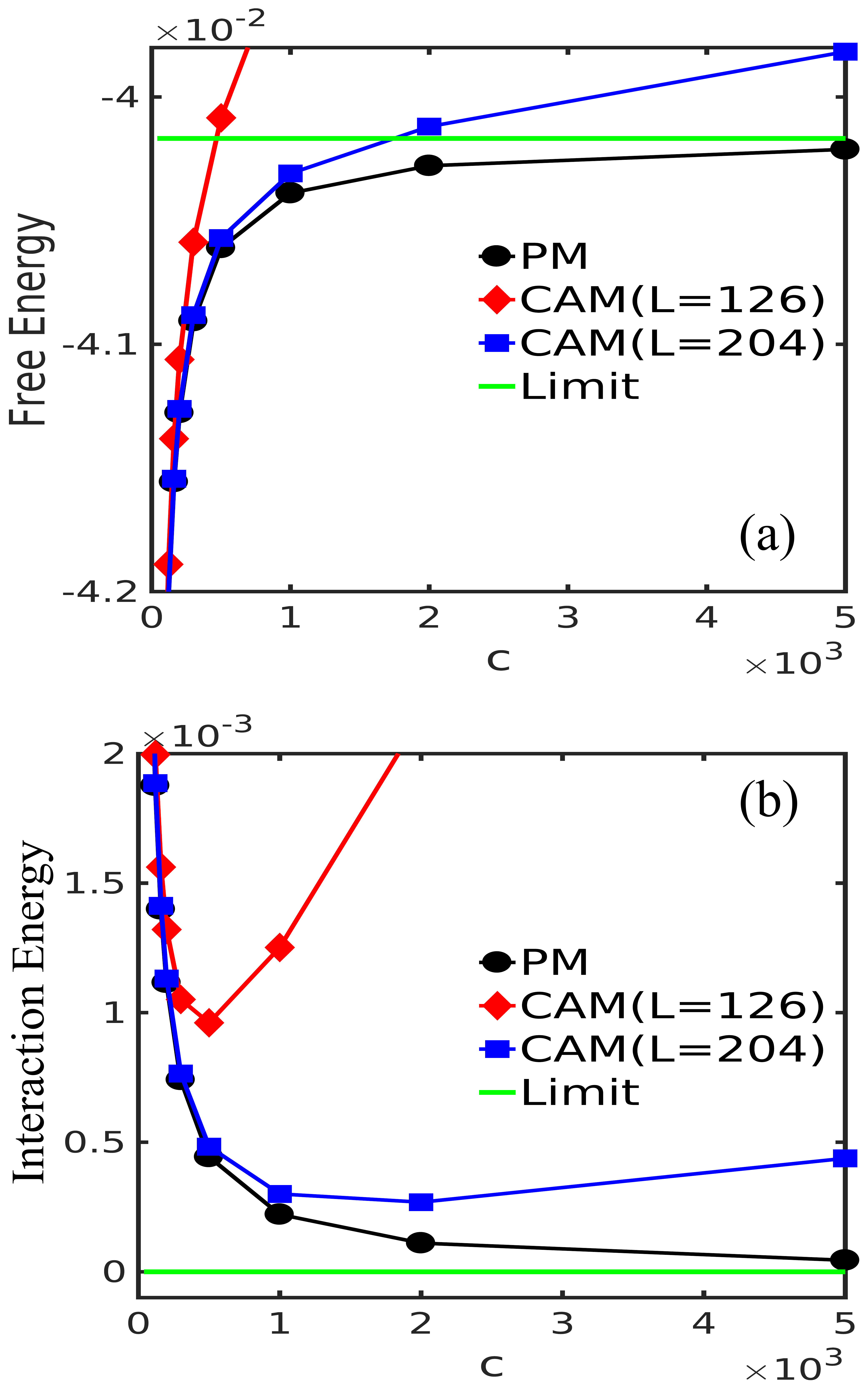}
	\caption{\label{fig.energy.PM.CAM}
		The parameters are set as $\tau=0$, $t=0$, $t_0=-0.3$, $g_0=-0.3$, $g_1=2.2$, \MC{$g_2=2.2$ and $ d_{0} = d_{1} = d_{2} = d_{3} = 0 $}.
		(a) Free energy calculated by PM and CAM, as a function of the penalty factor $c$, relative to that of hard constraint  ($c\to+\infty$).
		(b) Corresponding interaction energy computed by PM and CAM.
		}
\end{figure}
Subsequently, we will demonstrate the correctness of PM in evaluating free energy.
Subjected to the hard constraint $c\rightarrow\infty$, the nonzero Fourier modes are restricted on circles with radii $1$ and $q$. 
Under the symmetric assumption, we can analytically obtain the exact free energy function of any desired pattern, such as the DQC, by the TMAM\,\cite{chaikin1995principles, Lifshitz1997, Jiang2015, Jiang2017}. 
Theoretically, for an ordered structure, the value of free energy obtained by a correct numerical method should converge to that under the hard constraint case as $c$ increases.  
To observe this computational phenomenon, we apply PM and CAM with  $L=126$ ($E_{DA}=0.166879$), $L=204$ ($E_{DA}=0.091809$) to compute the DQC with same parameters and increasing penalty factor $c$ and observe the energy values. 
$16$ discrete points, corresponding to $16$ Fourier basis functions, in the length $2\pi$ of each direction are used in both approaches.
The green line denotes the free energy $F_{c\to\infty}$ under hard constraint whose energy expression is given by Eq.\eqref{eq.DQC.limit}.
It is obvious that the free energy is heavily dependent on $E_{DA}$. 
When the $E_{DA}$ decreases from $0.166879$ to $0.091809$, the intersection point where the free energy computed by the CAM nearly equals to $F_{c\to\infty}$ is from about $c=500$ to $c=1800$.
Therefore in order to obtain high accurate free energy by CAM, it is required to decrease the $E_{DA}$ with increasing computational area.
However, the free energy obtained by CAM is still divergent from $F_{c\to\infty}$ with the growth of $c$ as long as $E_{DA}$ does not vanish.  
In other words, CAM may not obtain the correct free energy when $c$ is large enough.
The essential reason is that the CAM cannot capture the real spectrum points as FIG.\,\ref{fig.pos.dif} shows.
Correspondingly the energy of interaction potential part \MC{diverges} quickly as $c$ becomes larger, as shown in FIG.\,\ref{fig.energy.PM.CAM} (b).
%#Correspondingly the energy of interaction potential part is divergent quickly as $c$ becomes larger, as shown in FIG.\,\ref{fig.energy.PM.CAM} (b).
On the contrary, the free energy and the interaction potential part obtained by the PM converge to the hard constraint case with the increase of $ c $.
Therefore, numerical results demonstrate that the PM can evaluate the free energy of quasicrystals correctly.

\begin{figure}[htbp]
	\centering
	\includegraphics[scale=0.15]{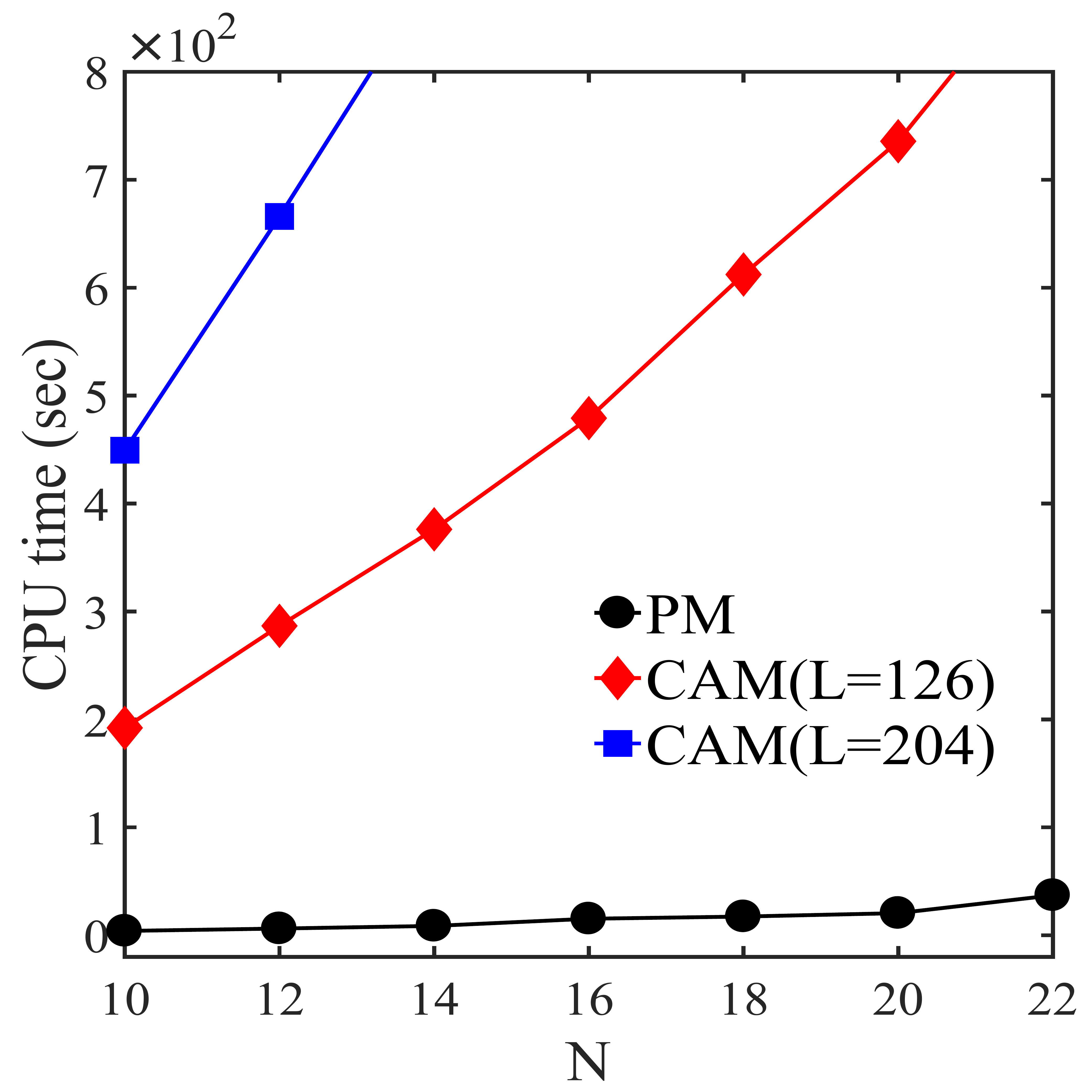}
	\caption{\label{fig.CPU}
		The CPU time of computing 2D DQCs by PM and CAM as a function of $N$ which is the number of discrete points in the length $2\pi$ of each direction.
		The black broken line represents the CPU time calculated by PM.
		The red and blue lines are both the CPU time by CAM but with $L=126$ and $L=204$, respectively.
		The model parameters are chosen as $c=80$, $\tau=0$, $t=0$, $t_0=-0.3$, $g_0=-0.3$, $g_1=2.2$, \MC{$g_2=2.2$ and $ d_{0} = d_{1} = d_{2} = d_{3} = 0 $}.
		}
\end{figure}
Furthermore, we compare the computational amount of two numerical methods in the above simulations.
FIG.\,\ref{fig.CPU} presents the CPU time of PM and CAM ($L=126$, $L=204$) with respect to different $ N $.
It is intelligible that the blue line ($ L = 204 $) is always above the red one ($ L = 126 $) due to the formidable growth of computational domain $ [0, L \cdot 2\pi)^2 $.
Meanwhile, $N$ is from $10 $ to $22$, PM (black line) costs much less computational amount than CAM regardless of $L=126$ or 204.
The reason is attributed to the implementation of PM in a high-dimensional unit cell $[0,2\pi)^{4}$.
The degree of freedom of PM is $ N^{4} $, while that of CAM is $\left( L \cdot N \right)^{2}$.
The computational cost of PM is much smaller than that of CAM, since $L$ grows faster than $N$.

\subsection{Candidate ordered phases}
\label{subsec.candidate}
Based on the PM, minimizing the free energy functional \eqref{eq.energy} of the CMSH model allows us to investigate the equilibrium phase behavior of quasicrystals and periodic crystals.
In what follows, we will focus on the occurrence and stability of 3D IQCs and related ordered structures, thus we will set the ratio between two characteristic length scales as $q=q_2/q_1=2\cos(\pi/5)$\,\cite{Subramanian2016, Jiang2017}.
Due to the choice of the ratio $q$, the 2D DQC is also considered as a possible equilibrium structure. 
In the PM, 3D IQCs are required to be embedded into $6$-dimensional periodic phases, and DQCs can be projected from the $4$-dimensional periodic structures.
The projection matrices have been given in our previous work\,\cite{Jiang2014, Jiang2015, Jiang2016, Jiang2017}.
In addition, a large number of related 2D periodic structures are also contained as the candidate structures in our study.
In the following simulations, the $n$-dimensional Fourier space is discretized by $16$ basis functions along each direction.
The total number of variables $ 16^n $ is enough to determine the relative stability of candidate phases according to the analysis in Sec.\,\ref{subsec.CAM.PM}.

\begin{figure}[htbp]
	\centering
	\includegraphics[scale=0.15]{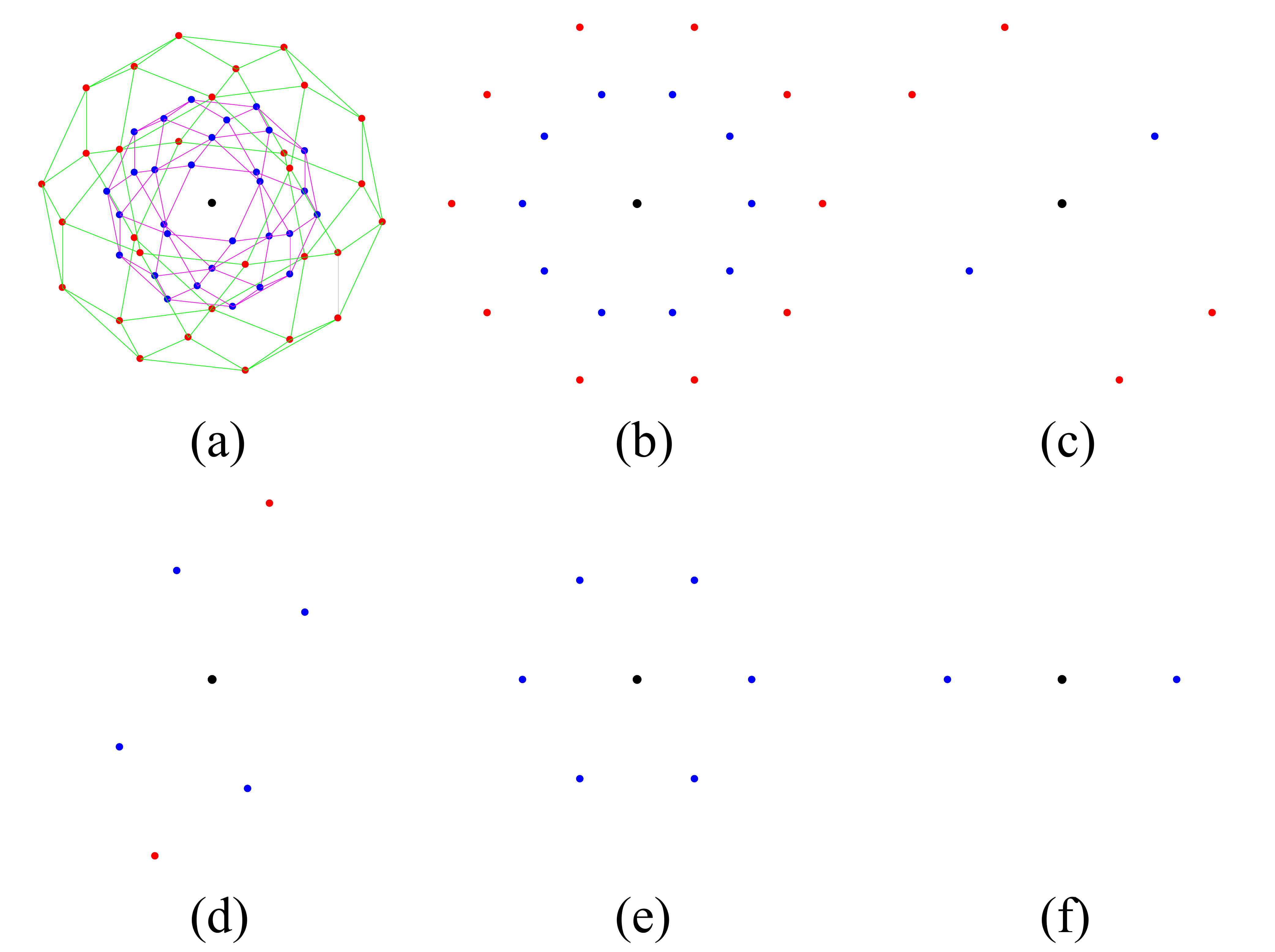}
	\caption{\label{fig.can.ini.spec}
		The initial distributions of Fourier wave vectors of candidate patterns with $q=2\cos(\pi/5)$.
		The central black dot represents the origin of the Fourier space.
		The blue and red dots are endpoints of wave vectors $\mathbf{k}_{\psi}$ and $\mathbf{k}_{\phi}$, respectively.
		(a)  A geometric figure, made by connecting the ends of $30$ basic modes on the spherical surface of radius $1$ and $q$ respectively, is two icosahedrons, with $20$ triangular faces and $12$ pentagonal faces.
		The initial configuration of the 2D DQC is given in (b) and that of the other periodic crystals in (c)-(f).
	}
\end{figure}

In order to quickly obtain desired ordered structures, the initial configurations are important in numerical simulations.
For the two-length-scales of the CMSH model, the initial nonzero Fourier vectors are chosen as FIG.\,\ref{fig.can.ini.spec} shows.
The central black dot represents the origin of the Fourier space and the blue (resp.~red) dots surrounding it are endpoints of wave vectors in $\mathbf{k}_{\psi}$ (resp.~$\mathbf{k}_{\phi}$).
Among these initial configurations, the first one is the IQC.
The blue (resp.~red) points represent $30$ diffraction points with icosahedral symmetry located on the spherical surface of radius $1$ (resp.~$q$).  
Edges on spherical surfaces with radii $1$ and $q$ are indicated by the magenta and luminous green lines, respectively.
For the other 2D patterns, the blue points stand for diffraction points located on the circle with radius $1$, while the red ones located on the circle with radius $q$.

\begin{figure}[htbp]
	\centering
    \includegraphics[scale=0.15]{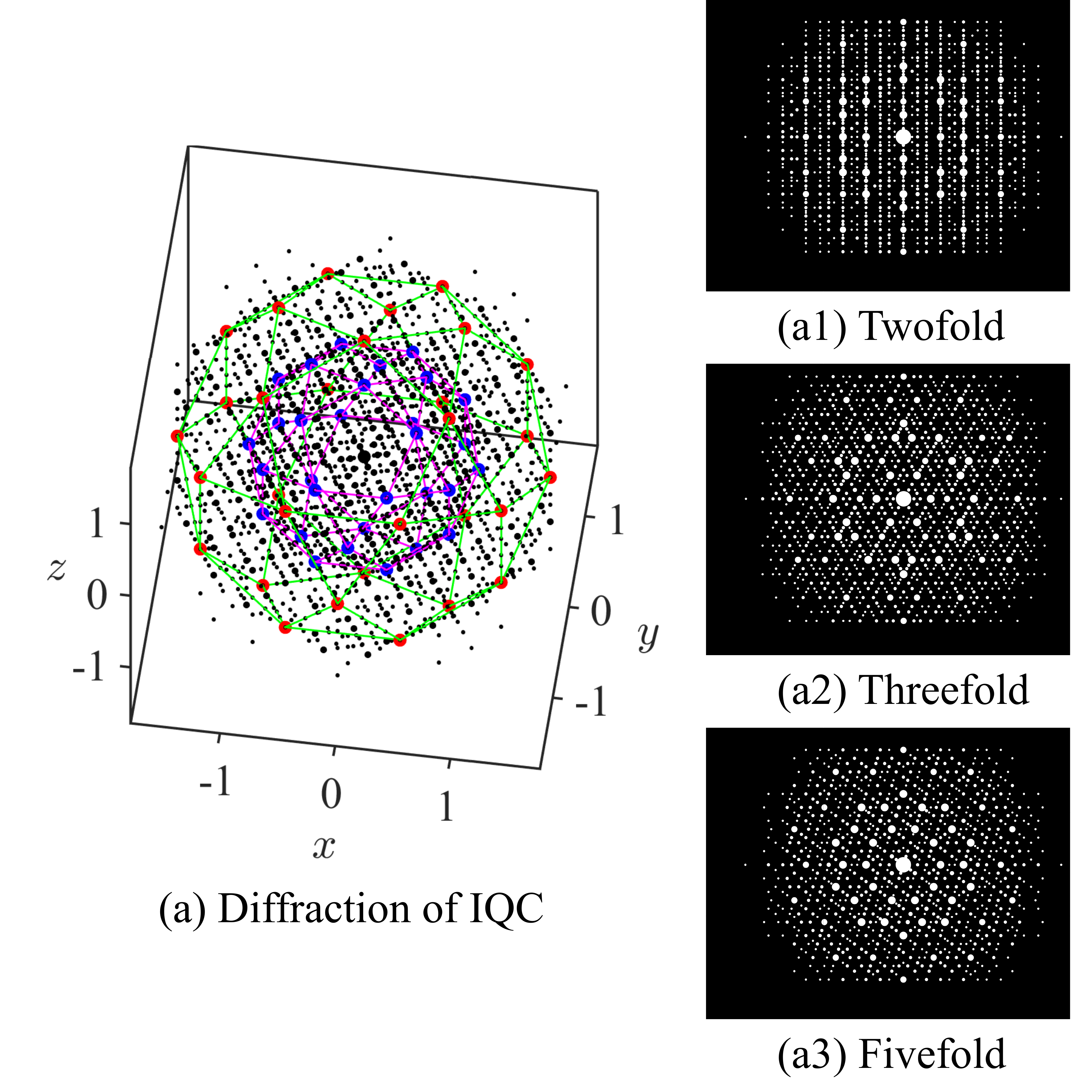}
    \includegraphics[scale=0.11]{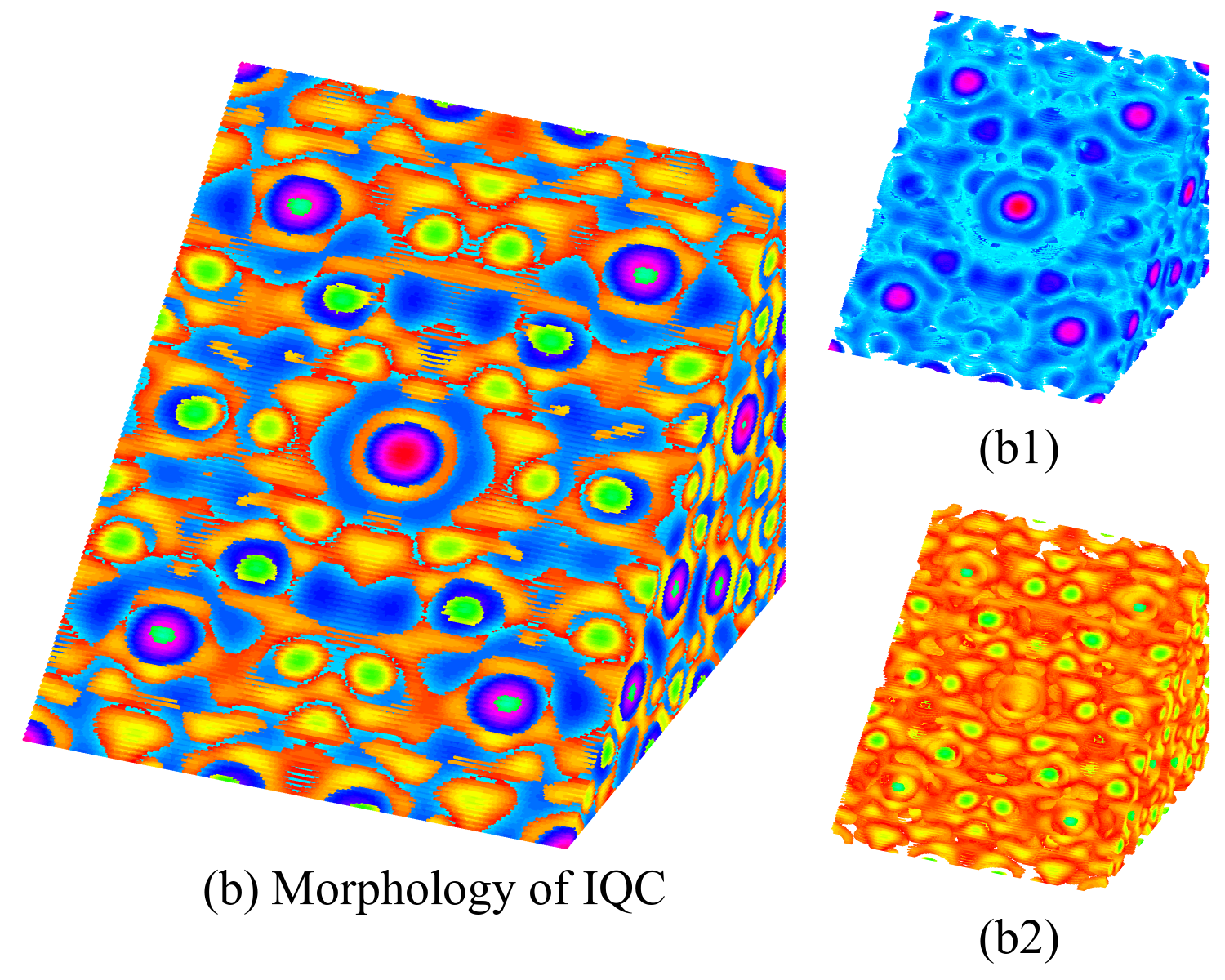}
	\caption{\label{fig.IQC}
		(a) The 3D diffraction of IQCs is computed by the PM with $q=2\cos(\pi/5)$.
		Only these Fourier modes whose diffraction intensity is larger than $1\times10^{-4}$ are shown. 
        The blue and red points are consistent with the initial configuration as shown in FIG.\,\ref{fig.can.ini.spec} (a).
		(a1)-(a3) Projecting the 3D diffraction pattern on 2D planes \MC{at different angles} exhibits twofold (a1), threefold (a2) and fivefold (a3) symmetric axes, respectively.
		%#(a1)-(a3) Projecting the 3D diffraction pattern on 2D planes by different angles exhibits twofold (a1), threefold (a2) and fivefold (a3) symmetric axes, respectively.
		(b) The physical space morphology of 3D IQCs. 
        The blue and orange colors present the dominant regions of components $\psi$ and $\phi$.
        The individual distributions of $\psi$ and $\phi$ are given in (b1) and (b2).
		}
\end{figure}
\begin{figure}[htbp]
	\centering
	\includegraphics[scale=0.15]{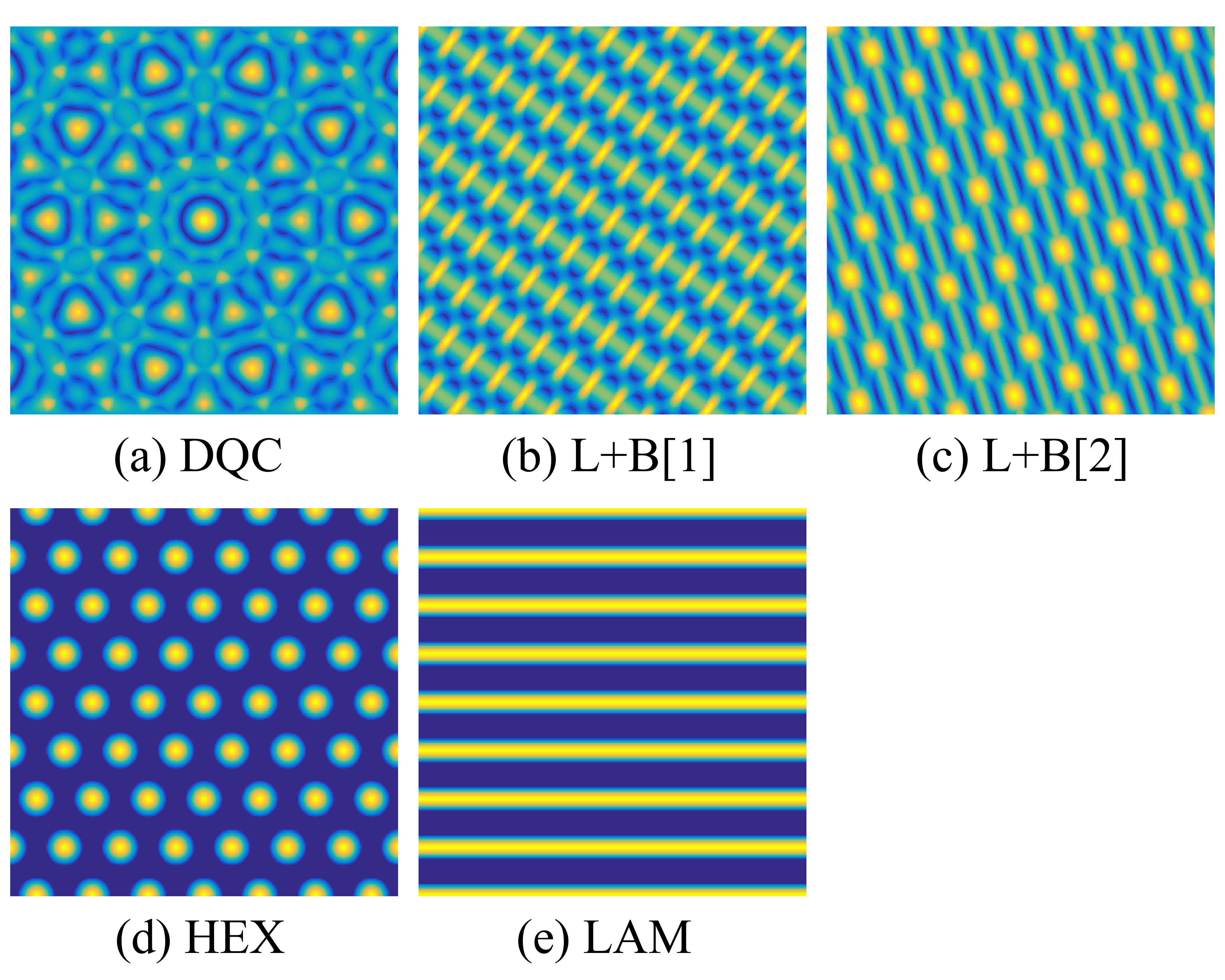}
	\caption{\label{fig.can.real}
        Morphologies of candidate ordered structures obtained by the PM using the initial values of FIG.\,\ref{fig.can.real}~(b)-(f) orderedly.
		(a) 2D DQCs.
		(b)-(c) L+B[1] and L+B[2] which are both the lamellar phases with alternating beads, but different density distribution.
		(d) Hexagonal phase (HEX).
		(e) Lamellar pattern (LAM).
	}
\end{figure}

Applying effective numerical methods and appropriate initial values outlined above to the dynamic equation \eqref{eq.AC}, we can obtain abundant equilibrium structures, as shown in FIG.\,\ref{fig.IQC} and FIG.\,\ref{fig.can.real}.
Among these phases, the diffraction patterns and the physical morphologies of 3D IQCs are given in FIG.\,\ref{fig.IQC}, whose initial nonzero Fourier vectors are illustrated with FIG.\,\ref{fig.can.ini.spec} (a).
These nonzero Fourier vectors with diffraction intensity larger than $1\times10^{-4}$ are given in FIG.\,\ref{fig.IQC} (a).
Except for a host of small black spots, the pattern is entirely consistent with the initial configuration of IQC in the Fourier space.
Projecting the 3D diffraction pattern of IQCs onto appropriate planes, we can found two-, three-, and five-fold symmetric axes in the spectra as shown in FIG.\,\ref{fig.IQC} (a1)-(a3).
These diffraction patterns are consistent with the experimental results of metallic alloys, such as Al-Cu-Fe, Zn-Mg-Sc, Al-Rh-Si alloys\,\cite{Tsai1987, Ishimasa1988, Kaneko2001, Koshikawa2003, Maezawa2004, Honma2007, Ishimasa2011}.
Correspondingly, FIG.\,\ref{fig.IQC} (b) displays its morphology by using the blue and orange colors to stand for components $ \psi $ and $ \phi $, respectively.
The individual morphologies of $\psi$ and $\phi$ are given in FIG.\,\ref{fig.IQC} (b1)-(b2).

The morphologies of other candidate phases are shown in FIG.\,\ref{fig.can.real} in which (a) gives 2D DQC, (b) and (c) present lamellar phase with alternating beads but different density distribution (L+B[1], L+B[2]), (d) and (e) are hexagonal (HEX) and lamellar (LAM) structures, respectively.
The initial non-zero Fourier vectors of these ordered patterns are given in FIG.\,\ref{fig.can.ini.spec}~(b)-(f) in sequence.

\subsection{Stability of IQCs and phase diagrams}
\label{subsec.results}
In the subsection, we will investigate the relative stability of these candidate ordered structures in two-component CMSH model under the hard constraint (the limiting case $c\to+\infty$) and the soft constraint (finite $c$).
Subjected to the hard constraint, the Fourier vectors of $\psi$ (resp.~$\phi$) should be strictly restricted on the circle with radius $1$ (resp.~$q$). 
Otherwise, the interaction energy becomes infinity.
As described in Sec.\,\ref{subsec.TMA}, for these candidate patterns whose basic wave vectors $\bfk$ can be found in FIG.\,\ref{fig.can.ini.spec}, their energy expressions are given in Eqs.\,\eqref{eq.IQC.limit}-\eqref{eq.HEX.limit} under the hard constraint. 
\begin{equation}
    \begin{aligned}
        F_{IQC,L} =\; &30\tau\widehat{\psi}^2 + 120g_0\widehat{\psi}^3 
        + 3330\widehat{\psi}^4 + 30t\widehat{\phi}^2 
        \\
        &+ 120t_0\widehat{\phi}^3 + 3330\widehat{\phi}^4 
        - 120g_1\widehat{\psi}^2\widehat{\phi} 
        \\
        &- 120g_2\widehat{\psi}\widehat{\phi}^2 \MC{ + 2100 d_{1}\widehat{\psi}^{2}\widehat{\phi}^{2} }
		\\
		& \MC{ + 1440 d_{2}\widehat{\psi}^{3}\widehat{\phi} + 1440 d_{3}\widehat{\psi}\widehat{\phi}^{3} },
    \end{aligned}
    \label{eq.IQC.limit}
\end{equation}
\begin{equation}
    \begin{aligned}
        F_{DQC,L} =\; &10\tau\widehat{\psi}^2 + 270\widehat{\psi}^4 
        + 10t\widehat{\phi}^2 + 270\widehat{\phi}^4 
        \\
        &- 20g_1\widehat{\psi}^2\widehat{\phi}
        - 20g_2\widehat{\psi}\widehat{\phi}^2 \MC{ + 140 d_{1}\widehat{\psi}^{2}\widehat{\phi}^{2} }
		\\
		& \MC{ + 60 d_{2}\widehat{\psi}^{3}\widehat{\phi} + 60 d_{3}\widehat{\psi}\widehat{\phi}^{3} },
    \end{aligned}
    \label{eq.DQC.limit}
\end{equation}
\begin{equation}
	\begin{aligned}
		F_{L+B[1],L} =\; & 2\tau\widehat{\psi}^2 + 6\widehat{\psi}^4 + 4t\widehat{\phi}^2 
		+ 36\widehat{\phi}^4 - 4g_2\widehat{\psi}\widehat{\phi}^2
		\\
		& \MC{ + 8 d_{1}\widehat{\psi}^{2}\widehat{\phi}^{2} },
	\end{aligned}
    \label{eq.LB1.limit}
\end{equation}
\begin{equation}
	\begin{aligned}
		F_{L+B[2],L} =\; & 4\tau\widehat{\psi}^2 + 36\widehat{\psi}^4 + 2t\widehat{\phi}^2 
		+ 6\widehat{\phi}^4 - 4g_1\widehat{\psi}^2\widehat{\phi}
		\\
		& \MC{ + 8 d_{1}\widehat{\psi}^{2}\widehat{\phi}^{2} },
	\end{aligned}
    \label{eq.LB2.limit}
\end{equation}
\begin{equation}
    F_{L2,L} =\; 2\tau\widehat{\psi}^2 + 6\widehat{\psi}^4,
    \label{eq.L2.limit}
\end{equation}
\begin{equation}
    F_{HEX,L} =\; 6\widehat{\psi}^2 + 90\widehat{\psi}^4,
    \label{eq.HEX.limit}
\end{equation}
where $\widehat{\psi},\widehat{\phi}\in\mathbb{R}$ stand for the Fourier coefficients of the order parameters $\psi$ and $\phi$, respectively.

\begin{figure}[htbp]
   \centering
	\includegraphics[scale=0.15]{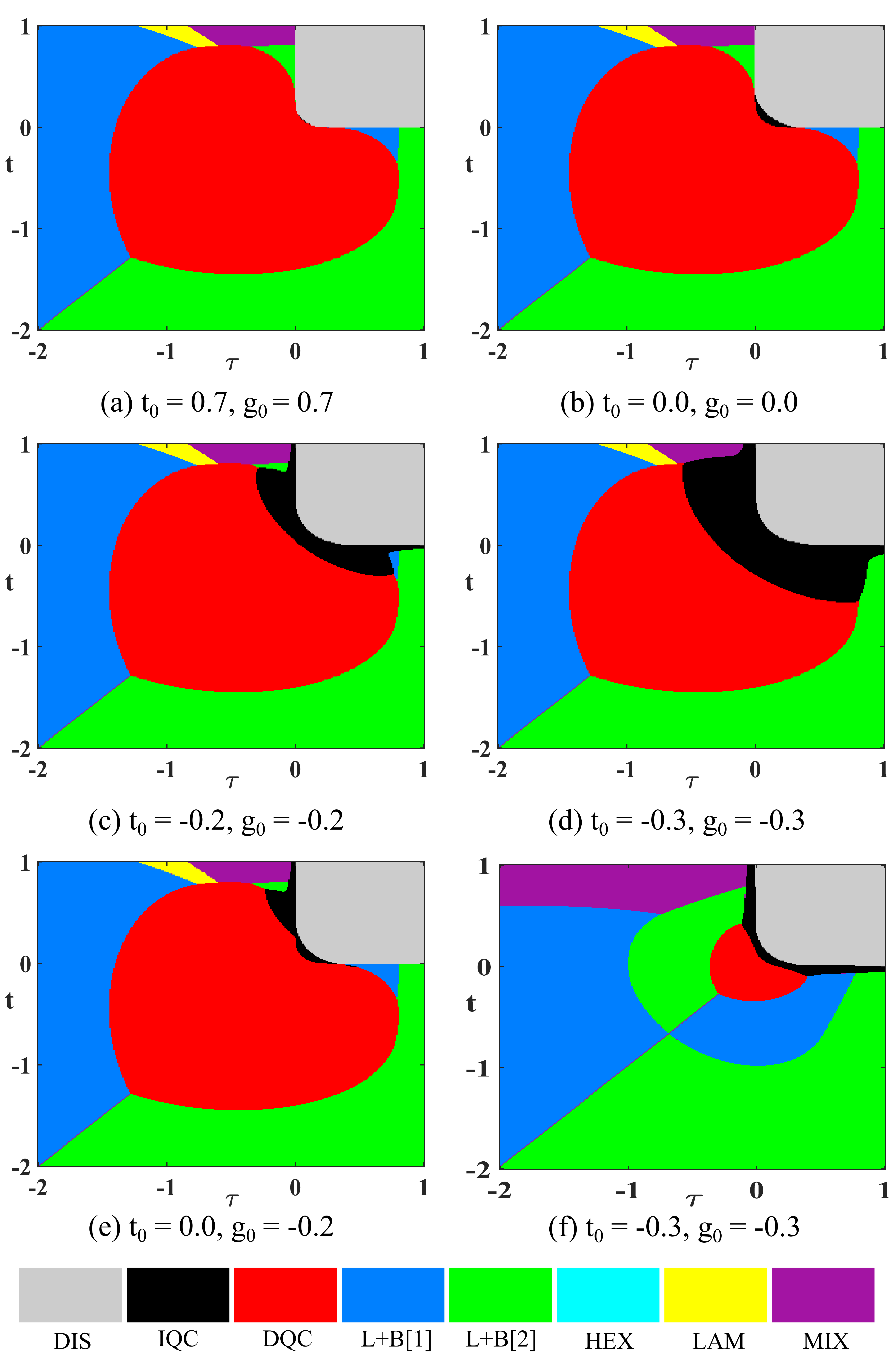}
	\caption{\label{fig.single.wave}
		$\tau$-$t$ phase diagrams with different parameters $t_0$ and $g_0$, but fixed $g_1=2.2$, $g_2=2.2$, \MC{(a)-(e) $ d_{0} = d_{1} = d_{2} = d_{3} = 0 $; (f) $ d_{0} = d_{1} = d_{2} = d_{3} = 1 $} under the hard constraint ($c\to+\infty$).
		In these phase diagrams, the different color represents different candidate structure.
		DIS is the disordered phase.
		The morphologies of 3D IQCs and other 2D ordered patterns can be found in FIG.\,\ref{fig.IQC} and FIG.\,\ref{fig.can.real}, respectively. 
		MIX denotes the region where the free energy of L+B[1] and LAM is not distinguishable.
		}
\end{figure}
From these analytical expressions, it is apparent that the parameters $t_0$ and $g_0$ only exist in the expression of IQCs.
It means that $t_0$ and $g_0$ can affect the relative stability of IQCs explicitly.
In order to analyze the influence, we first select different values of $t_0$, $g_0$, fix $q=2\cos(\pi/5)$, $g_1=2.2$, $g_2=2.2$, \MC{$ d_{0} = d_{1} = d_{2} = d_{3} = 0 $}, and leave $\tau$ and $t$ as free parameters to observe the phase behavior of these candidate patterns.
FIG.\,\ref{fig.single.wave} (a)-(e) present phase diagrams of $[t_0,g_0]=[0.7,0.7]$, $[t_0,g_0]=[0.0,0.0]$, $[t_0,g_0]=[-0.2,-0.2]$, $[t_0,g_0]=[-0.3,-0.3]$, and $[t_0,g_0]=[0.0,-0.2]$.
The stable region of the different candidate phase is denoted by the corresponding color.
MIX denotes the coexist region of L+B[1] and LAM in which the free energy of two phases is indistinguishable.
In FIG.\,\ref{fig.single.wave} (a), the stable area of IQCs in the phase diagram is almost invisible.
When $t_0=0$ and $g_0=0$, IQCs have an obvious stable region as shown in FIG.\,\ref{fig.single.wave} (b).
It means that the three-body interactions, corresponding to the cubic terms, can affect the stability of IQCs.
We further lower the values of $t_0$ and $g_0$, the stable region of IQCs grows rapidly, as FIG.\,\ref{fig.single.wave} (c)-(d) show.
It conducts that negative numbers of $t_0$ and $g_0$ are more beneficial to form and stabilize IQCs than the positive case.
The phase diagram with the unequal model parameters $t_0$ and $g_0$ is also given in FIG.\,\ref{fig.single.wave} (e).
Comparing with FIG.\,\ref{fig.single.wave} (b), we find that the black region extends to the part above the diagonal $\tau=t$ as $g_0$ descends.
On the contrary, FIG.\,\ref{fig.single.wave} (e) has less area of the black part under the diagonal than the figure (c) due to the growth of $t_0$.
As the model parameter $\tau$ or $t$ becomes negative, the absolute value of the cubic term becomes greater and leads to stronger three-body interactions.
Since IQCs have the largest number of the basic Fourier wave vectors, three-body interactions make more contributions to IQCs than the other candidate phases.
\MC{Then we investigate the influence of the last four terms through setting $ t_{0} = -0.3, g_{0} = -0.3, q = 2\cos(\pi/5), g_{1} = 2.2, g_{2} = 2.2, d_{0} = d_{1} = d_{2} = d_{3} = 1 $, and still letting $ \tau $ and $ t $ be free parameters to construct the phase diagram, as shown in FIG.\,\ref{fig.single.wave} (f).
Comparing with the subfigure (d), the stable regions of DQCs and IQCs dwindle sharply and LAM is invisible in the phase diagram.
It shows that the four-body interactions can affect the stability of quasicrystals as well as periodic crystals.
}

\begin{figure}[htbp]
	\centering
	\includegraphics[scale=0.15]{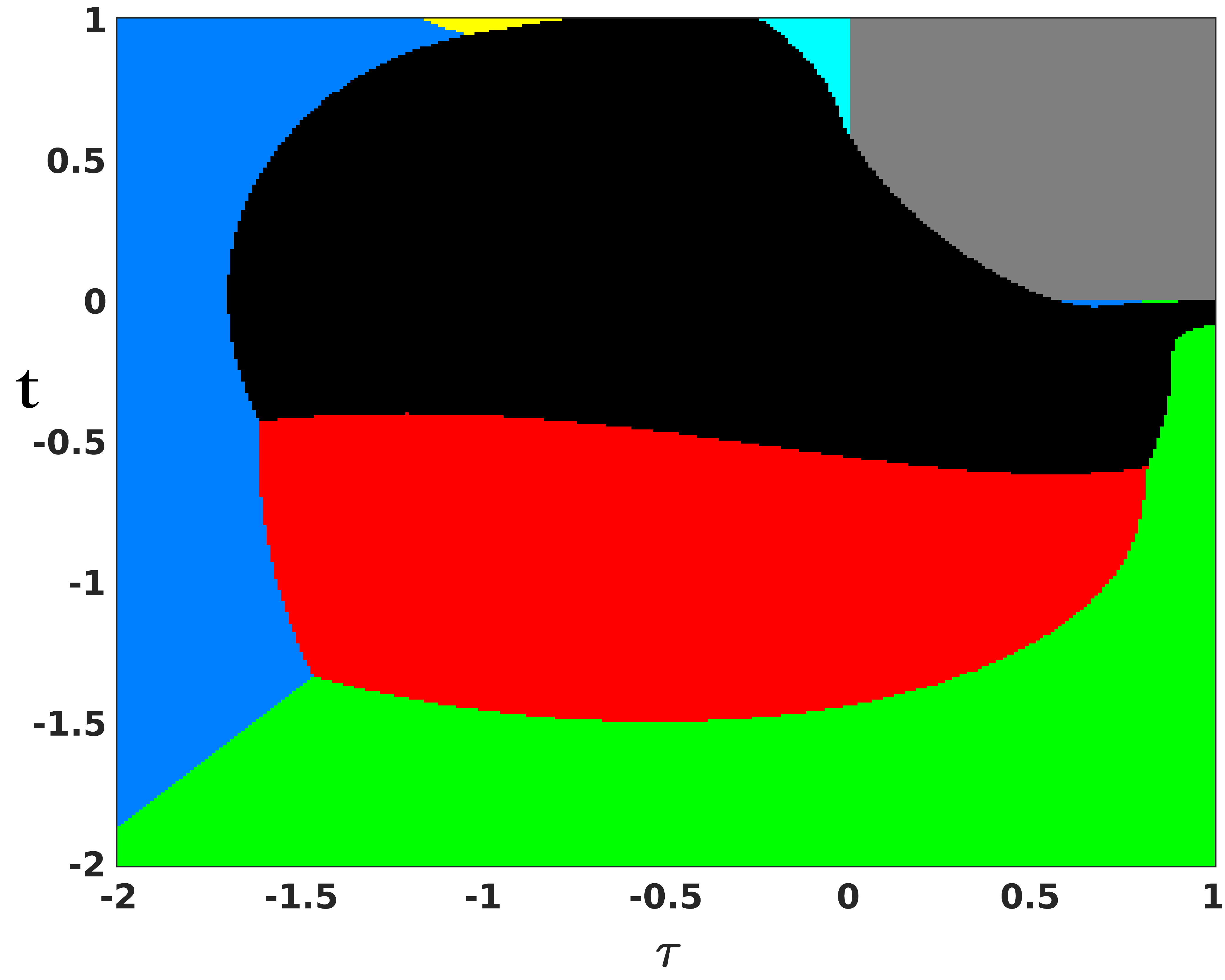}
	\caption{\label{fig.finite}
		The phase diagram under the soft constraint with $c=80$, $q=2\cos(\pi/5)$, $t_0=-0.3$, $g_0=-0.3$, $g_1=2.2$, \MC{$g_2 = 2.2$ and $ d_{0} = d_{1} = d_{2} = d_{3} = 0 $}.
		The colors have the same meaning \MC{as in FIG.\,\ref{fig.single.wave}.}
		%#The colors have the same meaning as FIG.\,\ref{fig.single.wave}.
		}
\end{figure}
We have explored the phase behavior of the two-component \MC{system described by the CMSH} model under the hard constraint with different parameters.
%#We have explored the phase behavior of the two-component system by the CMSH model under the hard constraint with different parameters.
Subsequently, we turn to study the relative stability of the system under the soft constraint (finite $c$) using the highly accurate PM.
Due to the expansive computational cost, we select model parameters, $c=80$, $q=2\cos(\pi/5)$, $t_0=-0.3$, $g_0=-0.3$, $g_1=2.2$, \MC{$g_2 = 2.2$ and $ d_{0} = d_{1} = d_{2} = d_{3} = 0 $}, as an example.
The parameters are the same as the hard constraint case of FIG.\,\ref{fig.single.wave} (d) except for the constraint factor $c$.
The corresponding phase diagram ($c=80$) is given in FIG.\,\ref{fig.finite}. 
It is obvious that the phase diagram of $c=80$ is very different from the hard constraint case of FIG.\,\ref{fig.single.wave} (d).  
The discrepancies can be ascribed to more complex behaviors of interaction terms in multi-component systems.
As shown in FIG.\,\ref{fig.finite}, DQC and L+B[1] occupy relatively large areas which are almost equal to the stable region of IQCs, but visibly smaller than the hard constraint case.
Under the soft constraint, the relative stability of IQCs has been significantly intensified and its stable region almost doubled in size. 
The reason is mainly attributed to the appearance of high-order non-zero Fourier modes which can form more cubic interactions to stabilize IQCs. Accordingly, the regions of DQC and L+B[1] have to shrink in the phase diagram. 
The phenomenon is different from the single component system in which the 2D DQC becomes metastable under soft constraint\,\cite{Jiang2017}.  
At the same time, the stable area of the L+B[2] phase remains \MC{unchanged}.
%#At the same time, the stable area of the L+B[2] phase remains unchangeable.
The HEX phase appears in the phase diagram and replaces the MIX region in FIG.\,\ref{fig.single.wave} (d).

The biggest distinction between the soft and the hard constraints is the emergence of high-order Fourier modes when $c$ is finite which do not locate on the circles with radii $1$ and $q$. 
To further investigate the phase behavior, it is useful to analyze the contribution to the free energy from different Fourier modes.
We split the Fourier modes into two parts: fundamental modes and higher-harmonics.
The fundamental modes are located on the circles with radii $1$ and $q$, as shown in FIG.\,\ref{fig.can.ini.spec}, and the higher-harmonics are the other Fourier modes.
The fundamental part of energy is defined by the contribution of the fundamental Fourier modes to the free energy, while the higher-harmonic energy is the remainder when subtracting the fundamental part from the total energy.

\begin{figure}[htbp]
	\centering
	\includegraphics[scale=0.15]{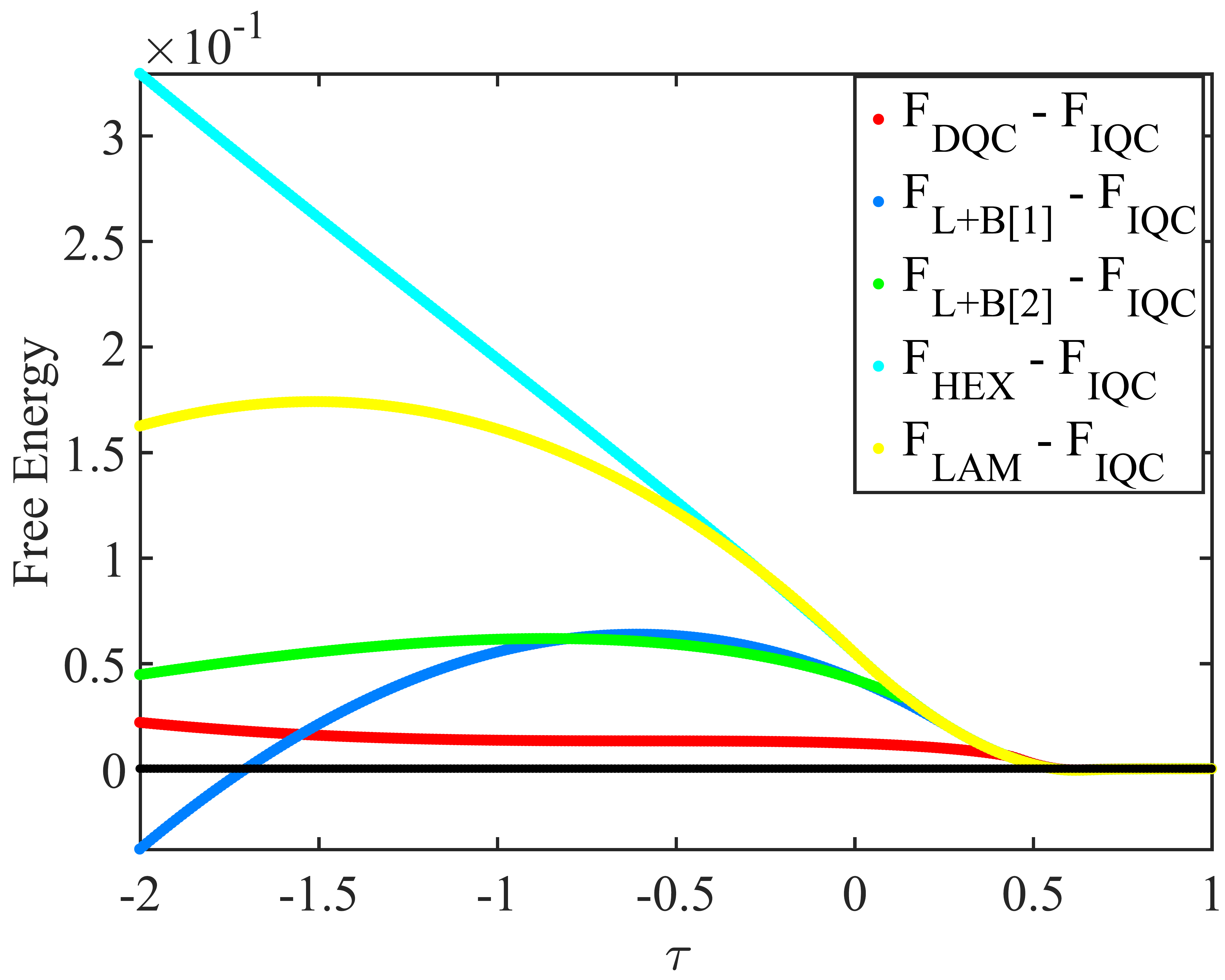}
	\caption{\label{fig.energy}
		The difference of the free energy of these candidate ordered structures from that of IQC as a function of $\tau$ along the phase path of fixed $c=80$, $q=2\cos(\pi/5)$, $t=0$, $t_0=-0.3$, $g_0=-0.3$, $g_1=2.2$, \MC{$g_2=2.2$ and $ d_{0} = d_{1} = d_{2} = d_{3} = 0 $}.
		}
\end{figure}

We take the line $t=0$ in the phase diagram of FIG.\,\ref{fig.finite} as an example to analyze the contribution to the free energy from the fundamental and higher-harmonic parts.
FIG.\,\ref{fig.energy} gives the free energy curves of these candidate phases.
To observe the tendency of the free energy better, we use the free energy of the IQC phase as the baseline.
As FIG.\,\ref{fig.energy} shows, the IQC is the most stable phase among these candidate ordered structures when $-1.70 \leqslant \tau \leqslant 0.5$.
While the L+B[1] phase is favored in the small range $ -2 \leqslant \tau \leqslant -1.71$.
It is consistent with the phase diagram of FIG.\,\ref{fig.finite}.

\begin{figure}[htbp]
	\centering
	\includegraphics[scale=0.15]{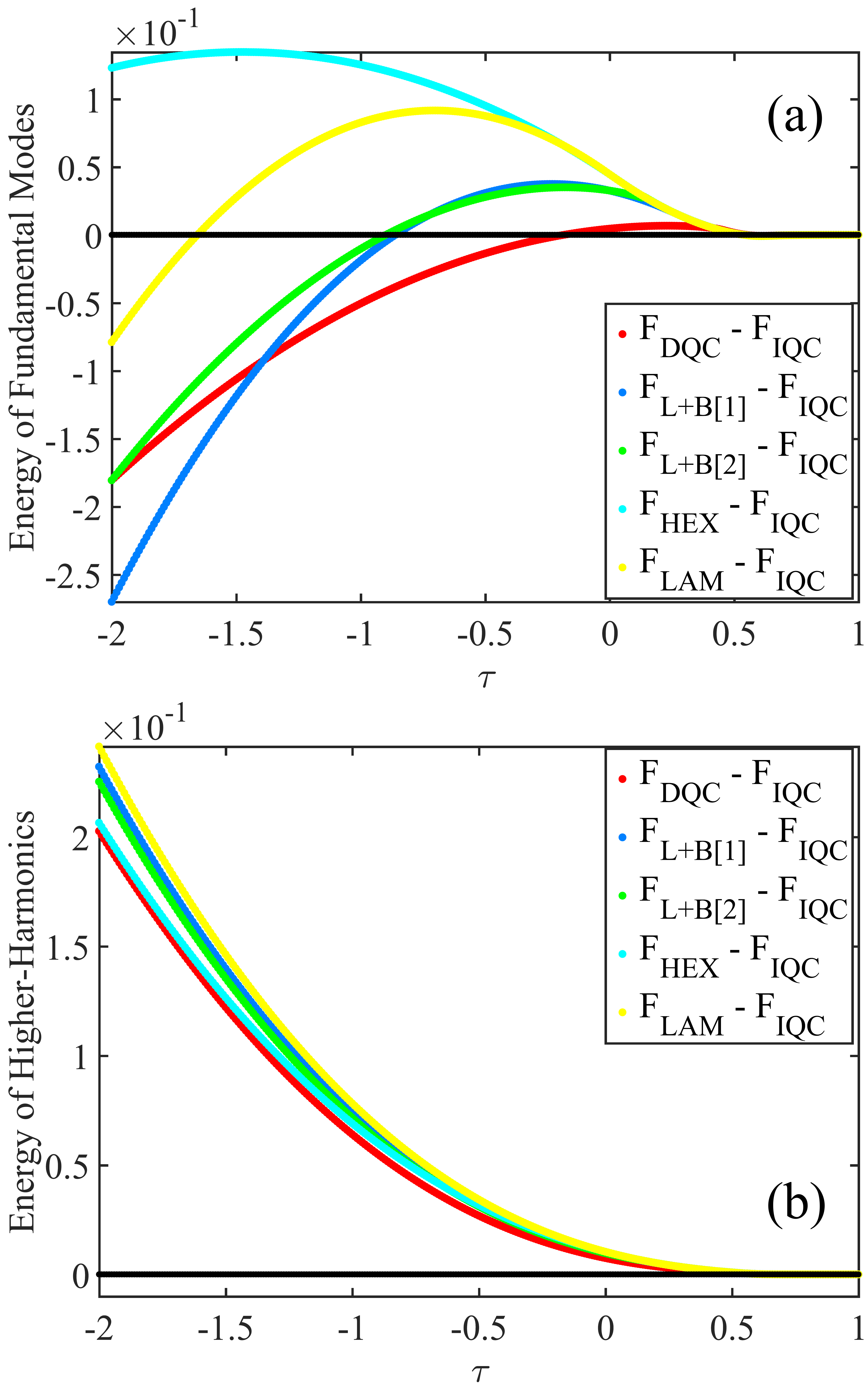}
	\caption{\label{fig.division.energy}
		The difference of (a) fundamental modes energy and (b) higher-harmonic energy of various candidate patterns from the corresponding part of IQCs as a function of $\tau$ for fixed $c=80$, $q=2\cos(\pi/5)$, $t=0$, $t_0=-0.3$, $g_0=-0.3$, $g_1=2.2$, \MC{$g_2=2.2$ and $ d_{0} = d_{1} = d_{2} = d_{3} = 0 $}.
		}
\end{figure}

We split the free energy ($t=0$) into the fundamental part and the higher-harmonic part, as FIG.\,\ref{fig.division.energy} presents.
The corresponding parts of IQCs are still utilized as baselines.
Firstly, FIG.\,\ref{fig.division.energy}~(a) gives the energy curves of fundamental modes of these candidate patterns.
The IQC phase has the largest number, $60$, of nonzero Fourier modes located on the spherical surfaces with radii $1$ and $q$, however, its fundamental energy is not the lowest among these phases in our considered region $\tau \leqslant 0.5$.
The reason is attributed to the fact that the quartic terms of IQC, which increase the value of free energy, are much larger than that of other candidate patterns as Eqs.\eqref{eq.IQC.limit}-\eqref{eq.HEX.limit} express.
Among these 2D ordered structures, the DQC is favored by the fundamental modes since it has the larger number, $20$, of nonzero Fourier modes located on the circles with radii $1$ and $q$, thus forming more triangles in the Fourier space and obtaining a lower fundamental energy, as shown in FIG.\,\ref{fig.division.energy} (a).
Meanwhile, there is a big difference between the L+B[1] and L+B[2] phases in the fundamental energy although they both have the same number of Fourier modes with nonzero coefficients.
From Eqs.\eqref{eq.LB1.limit}-\eqref{eq.LB2.limit}, we find that the analytical expression of L+B[2] can be obtained from the expression of L+B[1] by exchanging $\psi$ and $\phi$ if we treat $\tau$ and $t$ equally without discrimination.
It means that the components $\psi$ and $\phi$ play a completely contrary role in two different analytical expressions.
More precisely, the L+B[1] phase has more four-body interactions with respect to the component $\phi$ in the fundamental part than the L+B[2] structure does.
Based on the fact, we conclude that the component $\phi$ takes a more dominant position than $\psi$ to stabilize these candidate ordered structures with the given model parameters.

On the other hand, since the energy penalty factor $c$ is finite, the higher-harmonics cannot be ignored which have a profound impact on the free energy of the multi-component system, as FIG.\,\ref{fig.division.energy} (b) shows.
The IQC structure has the highest higher-harmonics contribution to the free energy, since its Fourier modes with nonzero coefficients form a large number of triad interactions that dramatically reduce the free energy.  
In contrast, the higher-harmonics contributions of 2D ordered structures generate fewer numbers of cubic interactions than that of the IQC does.
Therefore, the IQC has the lowest higher-harmonic energy.
Due to the competition of the fundamental modes and higher-harmonics, we obtain the stable regions of L+B[1] and IQC as shown in FIG.\,\ref{fig.finite}.

\section{SUMMARY}
\label{sec.summary}
In summary, we have investigated the thermodynamic stability of 3D IQCs in multi-component systems using a CMSH model with multi-length-scales. 
In the model system, the characteristic length scales contained in the interaction potential functions are acting on different order parameters through the positive-definite gradient terms. 
For two-component systems, when the ratio of the two-length-scales is set as the golden ratio, we predicted that the 3D IQCs and 2D DQCs are both stable for hard and soft constraints.
Under the hard constraint ($c\rightarrow\infty$), the Fourier modes are strictly restricted on the circles of radii $1$ and $q$. 
Using the two modes approximation method, we systematically
analyze the rule of emergence and stability of quasicrystals and a series of periodic crystals. 
Under the soft constraint (finite $c$), more higher-order non-zero Fourier modes arise besides those on the circles $|\mathbf{k}|=1$ and $|\mathbf{k}|=q$ which have a profound impact on the phase behavior. 
Thanks to the high-precision PM approach, it enables us to quantitatively analyze the phenomenon. 
Using the PM, high accurate free energy and more precise phase boundary have been obtained under the soft constraint.
Meanwhile, the advantages of PM over the traditional CAM have been also discussed in detail. 
%#\MC{In particular, the diffraction and morphology of the 3D IQC are consistent with the structures from lots of experimental observations in metallic alloys\,\cite{Tsai1987, Ishimasa1988, Kaneko2001, Koshikawa2003, Maezawa2004, Honma2007, Ishimasa2011} and that demonstrates the reliability of the PM.}
In particular, the numerical results about 3D IQCs are \MC{phenomenologically} consistent with lots of experimental observations in metallic alloys\,\cite{Tsai1987, Ishimasa1988, Kaneko2001, Koshikawa2003, Maezawa2004, Honma2007, Ishimasa2011} and that demonstrates the reliability of the PM.
These results provide a good understanding of the stability of 3D IQCs in multi-component systems.
The numerical approaches and insights will be helpful for further studying related quasiperiodic physical systems.

\section*{Acknowledgements} 
This work is supported by the National Natural Science Foundation of China (11771368), Hunan Science Foundation of China (2018JJ2376),
Youth Project Hunan Provincial Education Department of China (Grant No. 16B257), and Project of Scientific Research Fund of Hunan Provincial Science and Technology Department (2018WK4006).

\begin{appendix}

\section{CAM}
\label{app.introd.CAM}
The principal idea of CAM is using the periodic structures to approximate the quasiperiodic structures.
For a $d$-dimensional quasicrystal, its reciprocal lattice vector $\bfk$ can be expressed by $d$ linearly independent reciprocal vectors, $\bfe^{*}_1, \bfe^{*}_2, \cdots, \bfe^{*}_d$,
\begin{equation}
    \bfk = p_1\bfe^{*}_1 + p_2\bfe^{*}_2 + \cdots + p_d\bfe^{*}_d.
\end{equation}
It is important to note that $\bfk$ cannot be represented by linear combinations of $\bfe^{*}_{i}$ with integer-valued coefficients since there always exists, at least, an irrational number $p_i\in\mathbb{R}$.
However, the quasiperiodic function $\phi(\bfr)$ can be approximately expanded as
\begin{equation}
    \phi(\bfr) = \sum_{\bfk} \hphi(\bfk) e^{i\cdot(L\bfk)\cdot\bfr/L},
    ~~~~~ \bfr\in[0,2\pi L)^d.
\end{equation}
If there exists a rational number $L$ such that $Lp_i\in\mathbb{Z}$ or $Lp_i$ can be sufficiently close to a series of integers, then
\begin{equation}
    \left|L\cdot(p_1,\cdots,p_d) - ([Lp_1],\cdots,[Lp_d])\right|_{l^\infty} 
    \to 0,
    \label{eq.CAM}
\end{equation}
where $[y]$ represents the nearest integer of the real number $y$.
Replacing $[L\bfk]$ by $L\bfk$, numerical methods designed for periodic structures can be used to treat quasicrystals.
It should be noted that the rational number $L$ depends on these irrational numbers $p_i$ due to rotational symmetry and the desired precision of the approximation.
Without loss of generality, we can always choose $(1,0,\cdots,0)$ as one of the primitive reciprocal vectors which leads $L$ to be an integer.

\begin{figure}[htbp]
    \centering
    \includegraphics[scale=0.15]{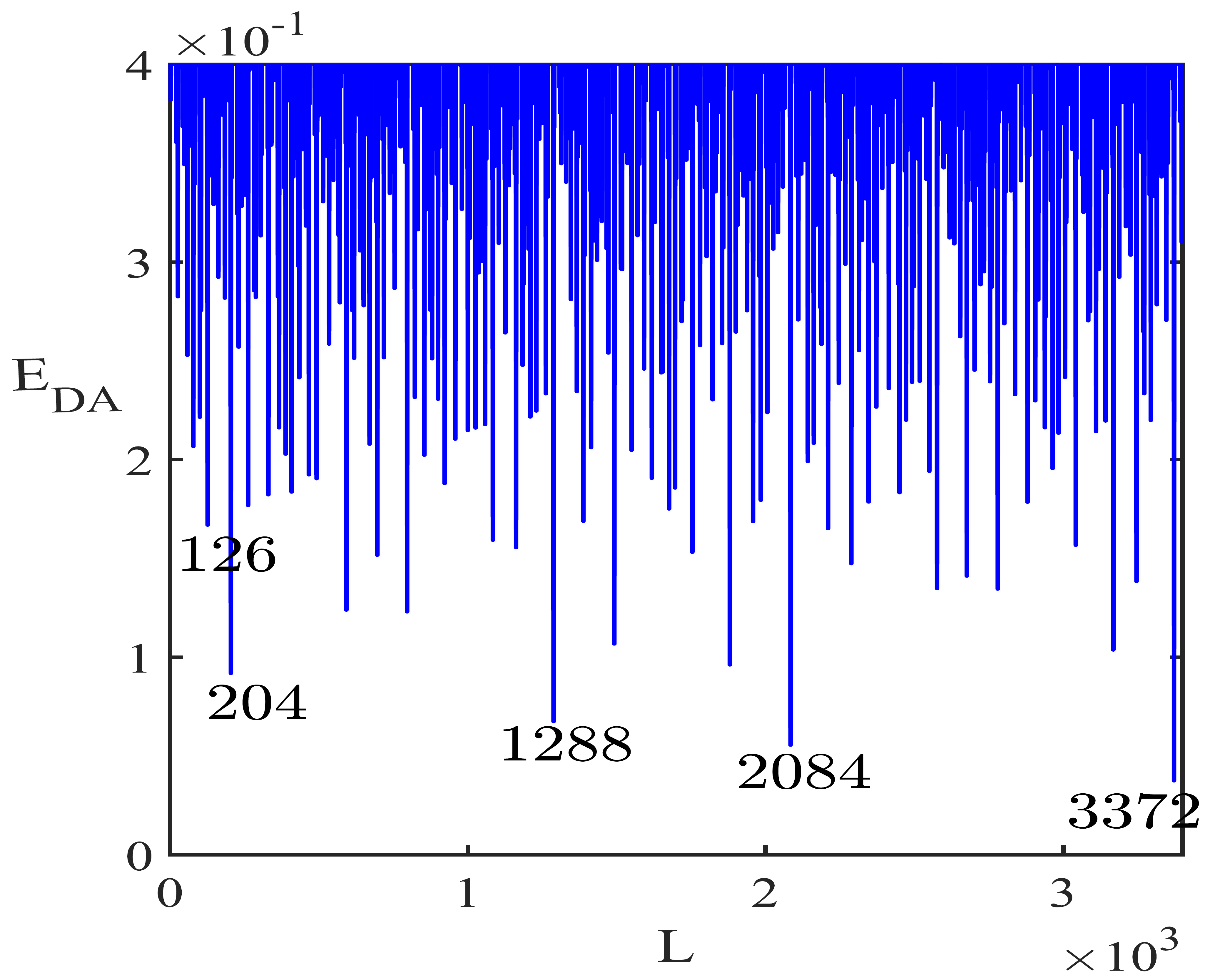}
    \caption{\label{fig.DA}
		The DA error $E_{DA}$ as a function of $L$ for the decagonal quasicrystalline order.
    }
\end{figure}

\begin{figure}[htbp]
    \centering
    \includegraphics[scale=0.15]{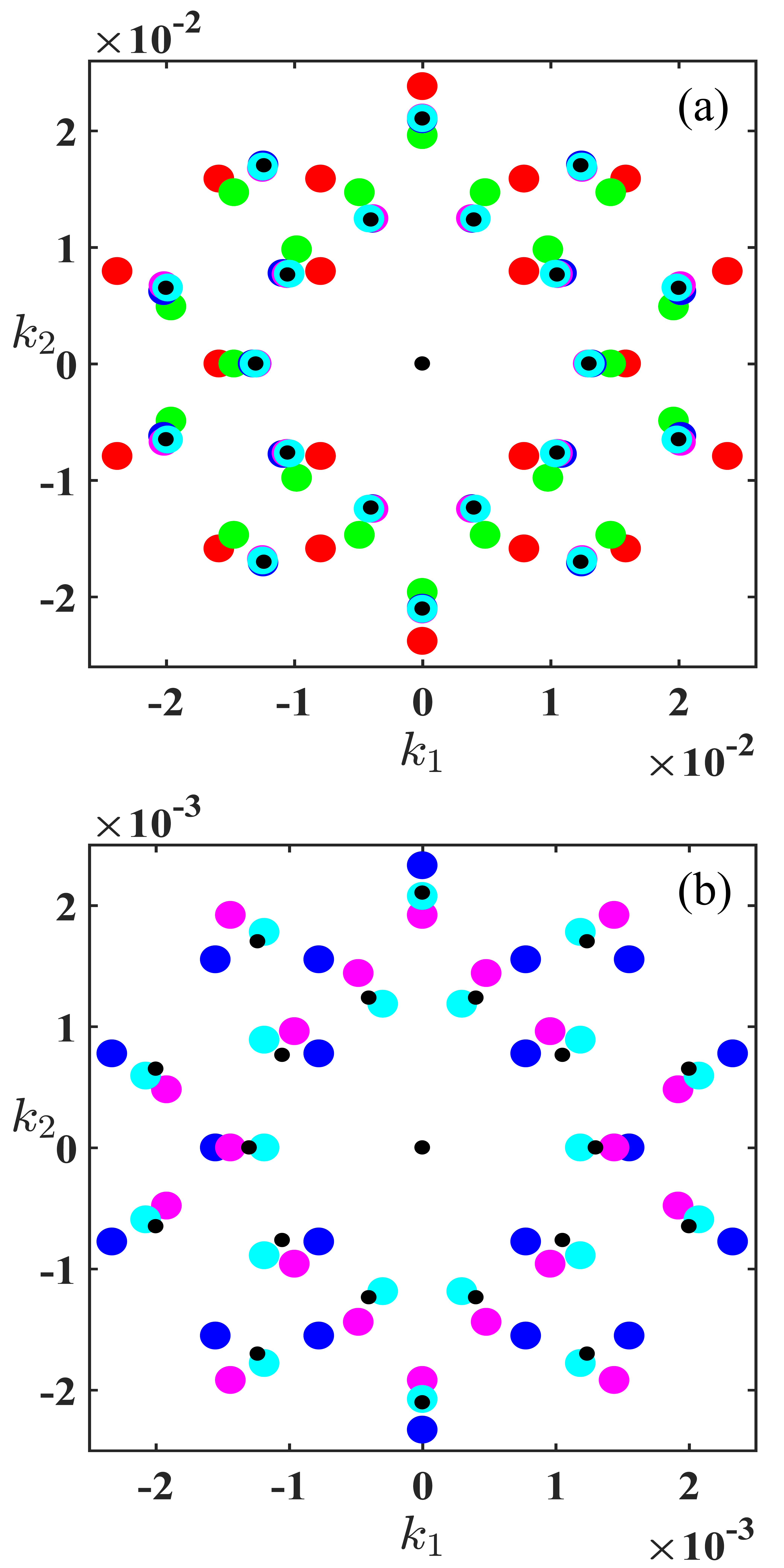}
    \caption{\label{fig.pos.dif}
        (a) The approximation and authentic positions of the largest Fourier spectrum points of the DQCs.
        (b) The magnified counterpart of the pattern (a).
        The central black dot is the origin and the surrounding black points represent the actual positions.
        The red, green, blue, magenta and light blue points give the approximate positions with $L=126$, $204$, $1288$, $2084$ and $3372$, respectively.
        }
\end{figure}

How to determine $L$ is the so-called \textit{Diophantine Approximation} (DA) topic in the number theory which deals with the approximation of real numbers by rational numbers or integers.
In FIG.\,\ref{fig.DA}, we give DA error, $E_{DA}$, as a function of the integer $L$ with the decagonal quasicrystalline order.
According to our experiments, the DQCs can be computed by CAM at least $L=126$.
We mark the first integer $126$ and the other integers which have lower $E_{DA}$ in FIG.\,\ref{fig.DA}.
These marked integers show that $L$ increases very quickly as the desired precision improves slightly (also see TAB.\,\ref{tab.DA}).
Then we give the approximation and actual positions of these Fourier spectrum points to analyze the influence of $E_{DA}$, as shown in FIG.\,\ref{fig.pos.dif}.
The central black point is the origin of coordinates and its surrounding black points represent the actual positions.
The red, green, blue, magenta and light blue dots show the approximated positions with $L=126$, $204$, $1288$, $2084$ and $3372$, respectively.
It is obvious that the green points, $L=204$, are significantly closer to the real positions than the red ones which correspond to $L=126$.
Although the positions of the \MC{points with the remaining colors can hardly be distinguished from} the authentic ones in FIG.\,\ref{fig.pos.dif} (a), \MC{a gap} still exists if we zoom in this figure, as shown in FIG.\,\ref{fig.pos.dif} (b).
%#Although the positions of the points with the rest colors hardly distinguish with the authentic ones in FIG.\,\ref{fig.pos.dif} (a), the gap still exists if we zoom in this figure, as shown in FIG.\,\ref{fig.pos.dif} (b).
As a matter of fact, the $E_{DA}$ always exists in CAM unless $L\to+\infty$.

\end{appendix}

\end{document}